\documentclass[twoside,11pt]{article}

\usepackage{algorithm}
\usepackage{algpseudocode}
\usepackage[preprint]{jmlr2e}
\usepackage{amsmath}
\usepackage{mathtools}
\usepackage{booktabs}
\usepackage{tabularx}
\usepackage{multirow}
\usepackage{makecell}
\usepackage{adjustbox}
\usepackage{colortbl}
\usepackage{tikz}
\usepackage{enumitem}
\usepackage{xcolor}
\usepackage{microtype}
\usepackage{placeins}
\usetikzlibrary{arrows.meta,calc,positioning}

\hypersetup{hidelinks}

\newcommand{\qhimap}{\textsc{Q-HiMAP}}
\newcommand{\qaffine}{\textsc{Q-affine}}
\newcommand{\bc}{\boldsymbol{c}}
\newcommand{\bb}{\boldsymbol{b}}
\newcommand{\bell}{\boldsymbol{\ell}}
\newcommand{\bz}{\boldsymbol{z}}
\newcommand{\bg}{\boldsymbol{g}}
\newcommand{\bu}{\boldsymbol{u}}
\newcommand{\bv}{\boldsymbol{v}}
\newcommand{\bh}{\boldsymbol{h}}
\newcommand{\bx}{\boldsymbol{x}}

\definecolor{HiMAPBest}{RGB}{244,177,131}
\definecolor{HiMAPSecond}{RGB}{252,228,214}
\newcommand{\bestscore}[1]{\cellcolor{HiMAPBest}#1}
\newcommand{\secondscore}[1]{\cellcolor{HiMAPSecond}#1}
\jmlrheading{1}{2026}{1--1}{8/26}{--}{--}{Tao Wang, Qiannan Huang, Jun Zhu, and Cheng Meng}
\ShortHeadings{Hilbert Mass-Addressed Barycenters}{Wang et al.}
\firstpageno{1}

\begin{document}

\title{HiMAP: Hilbert Mass-Addressed Parameterization for Multivariate Barycenters and Fr\'echet Regression}

\author{
\name Tao Wang
\email wang\_tao@ruc.edu.cn\\
\addr Institute of Statistics and Big Data\\
Renmin University of China\\
59 Zhongguancun Street, Haidian District\\
Beijing 100872, China
\AND
\name Qiannan Huang
\email qnhuang@connect.hku.hk\\
\addr Area of Innovation and Information Management\\
Faculty of Business and Economics\\
The University of Hong Kong\\
Pokfulam Road, Hong Kong, China
\AND
\name Jun Zhu
\email dfsxzj@ruc.edu.cn\\
\addr Institute of Statistics and Big Data\\
Renmin University of China\\
59 Zhongguancun Street, Haidian District\\
Beijing 100872, China
\AND
\name Cheng Meng$^\dagger$
\email chengmeng@ruc.edu.cn\\
\addr Institute of Statistics and Big Data\\
Renmin University of China\\
59 Zhongguancun Street, Haidian District\\
Beijing 100872, China\\
$^\dagger$Corresponding author
}

\editor{Under review}

\maketitle

\begin{abstract}
Learning from multivariate distribution-valued data requires both averaging
distributions and predicting them from covariates.
Positive barycenters and Fr\'echet regression impose different closure
requirements.
Regression weights can be negative, yet every fitted value must remain a
probability law.
We introduce the Hilbert mass-addressed parameterization (HiMAP), which uses
balanced recursive median partitions to assign the same probability mass to
each binary address across distributions.
The common mass address yields two complementary representations.
Q-HiMAP averages physical paths and defines a closed-form barycenter for
nonnegative weights.
Tree-logit HiMAP (TL-HiMAP) maps root geometry and relative split positions to
Hilbert coordinates, where every finite affine combination decodes uniquely to
a probability law.
We establish exact mass coding, an encoder--decoder inverse, completeness,
Wasserstein continuity, dense model coverage, and deterministic error
decompositions for fixed-size particle outputs.
The TL coordinates also give closed-form global and local Fr\'echet regression
and allow each response to be encoded once for repeated prediction.
Experiments with certified barycenters, simulated distribution regression,
and two NHANES survey cycles show accurate barycenter recovery, algebraically
exact TL-coordinate composition with finite particle reconstruction error,
competitive predictive loss, and efficient repeated prediction.
\end{abstract}


\section{Introduction}
\label{sec:introduction}

Probability distributions arise as observations when each statistical unit is
represented by a sample or population rather than a single vector.
Examples include cell populations, sequential distributions, and repeated
measurements within participants
\citep{LinShiYeLi2023MAWBarycenterSingleCell,
ChengAeronHughesMiller2023DynamicalWassersteinBarycenters}.
Analyzing such data requires two basic operations.
One operation averages several distributions, and the other predicts a
distribution from covariates.
The Wasserstein metric preserves spatial information and multivariate
dependence, and its Fr\'echet mean defines the Wasserstein barycenter
\citep{AguehCarlier2011Barycenters,panaretos2019statistical,
peyre2019computational}.
Fr\'echet regression uses the same metric-space construction for conditional
prediction
\citep{PetersenMuller2019FrechetRegression}.

In one dimension, a quantile function separates a common probability label
from its distribution-specific spatial location.
For nonnegative weights, pointwise averages of quantile functions remain
monotone and give the Wasserstein barycenter.
This quantile geometry has supported Wasserstein regression,
distribution-on-distribution models, and autoregressive distribution models
\citep{chen2023wasserstein,ghodrati2022distribution,
zhu2023autoregressive}.
Signed affine combinations can lose monotonicity, however, because global and
local regression weights may be negative.
In several dimensions, no canonical scalar order has the full algebra of a
univariate quantile function.
Multivariate distribution regression must therefore align probability mass
across distributions while preserving probability laws under signed
combinations.

Existing multivariate barycenter methods follow several geometric routes.
Direct optimal transport solves or regularizes the Wasserstein objective using
linear programming, entropic iterations, or stochastic optimization
\citep{CuturiDoucet2014FastWassersteinBarycenters,cuturi2013sinkhorn,
GenevayCuturiPeyreBach2016StochasticOT}.
Projection and slicing methods reduce the problem to lower-dimensional
transport calculations
\citep{BonneelRabinPeyrePfister2015SlicedRadonBarycenters,
LinFanHoCuturiJordan2020ProjectionRobustWasserstein}.
Linearized optimal transport instead represents laws through transport maps
from a reference distribution
\citep{BigotKlein2018BarycenterOTMaps,
WerenskiMalleryAeronMurphy2025Linearized,
GunsiliusHsiehLee2024Tangential}.

Related representations attach spatial information to multivariate
probability ranks.
Vector quantiles, center-outward ranks, and entropy-regularized soft ranks use
transport maps or couplings
\citep{CarlierChernozhukovGalichon2017VectorQuantileRegression,
HallinDelBarrioCuestaAlbertosMatran2021CenterOutward,
MasudEtAl2023SoftRank}.
Quantile pyramids build recursive fixed-mass partitions, while Hilbert-curve
methods provide reproducible traversals of multivariate samples
\citep{HjortWalker2009QuantilePyramids,
AnMacEachern2024DependentQuantilePyramids,li2024hilbert}.
Together, these methods provide spatial coordinates, ranks, or distances for
multivariate distributions.
Our aim is to obtain a common probability address that supports both a spatial
barycenter and an affine coordinate representation of probability laws.

The choice of response geometry is also important for distribution regression.
Metric-space Fr\'echet regression supplies global, local linear, and adaptive
random-forest weights
\citep{PetersenMuller2019FrechetRegression,QiuYuZhu2024RandomForestFrechet}.
Conditional entropic barycenters and nonparanormal transport provide recent
multivariate implementations
\citep{FanMuller2025ConditionalWassersteinBarycenters,
ParkGaynanova2026NPTFrechet}.
Direct signed Wasserstein formulations study existence, uniqueness, and
duality for real barycentric coefficients
\citep{TornabeneVeneroniSavare2025Generalized,
JacobsZhou2026SignedWasserstein}.
Our coordinate approach seeks a representation that aligns equal masses,
admits closed-form positive averages, and preserves probability laws under
affine aggregation.
The representation should operate directly on point clouds and allow one response encoding to
serve multiple prediction targets.

We introduce the \emph{Hilbert mass-addressed parameterization}, abbreviated
HiMAP.
Starting from the smallest axis-aligned support rectangle, HiMAP recursively
splits each parent cell at a conditional median according to a fixed balanced
schedule.
Every binary address of depth $l$ then represents mass $2^{-l}$ for each
encoded law.
The common address supports two complementary representations.
Q-HiMAP averages the physical paths attached to each address, while Tree-logit
HiMAP (TL-HiMAP) applies a buffered logit transform to relative
split positions and combines them with root coordinates in a real Hilbert
space.
For nonnegative weights, the Q representation gives a spatial barycenter.
The TL representation also permits signed weights, returns a probability law,
and allows the decoded output to be encoded and combined again.

\textbf{Contributions.}
\begin{enumerate}[leftmargin=2.1em,label=(\roman*)]
\item \textbf{Mass-addressed representation.}
We prove exact mass coding, an encoder and decoder inverse, and completeness of
the TL space.
Transport comparison and continuity results connect both induced metrics
to Wasserstein distance.
The union of buffered HiMAP models is $W_2$-dense in the finite-second-moment
laws.
\item \textbf{Barycenters and affine closure.}
Convex combinations of Q-HiMAP path maps remain valid path maps and give
closed-form barycenters.
Affine combinations of TL coordinates remain in the TL coordinate space,
decode uniquely, and make population aggregation exactly compositional.
\item \textbf{Distribution regression and finite computation.}
Because TL coordinates are closed under affine combinations, global and local
Fr\'echet regression reduce to one Hilbert-valued average followed by one
decode.
The point-cloud algorithms return fixed-size particle laws.
Their deterministic error decompositions separate regression, encoding,
truncation, and integer allocation errors.
\end{enumerate}

The empirical study uses three independently certified positive barycenter
problems, global and local regression simulations, hierarchical composition,
repeated-target timing, and two NHANES survey cycles.
Wasserstein-based criteria and the native HiMAP geometries show accurate
certified barycenters, TL-coordinate composition to numerical precision, and
competitive distribution regression; finite particle reconstruction accounts
for the remaining output discrepancy.
Repeated predictions also benefit from reusing the encoded responses.

Figures~\ref{fig:himap-address-construction}
and~\ref{fig:himap-aggregation} show the recursive address construction and
the two aggregation rules, respectively.
Section~\ref{sec:preli} defines these objects, and
Section~\ref{sec:representation} establishes the representation theory,
barycentric closure, stability, model coverage, and population regression.
The finite particle algorithms, deterministic error decompositions, and
consistency results follow in Section~\ref{sec:estimation}.
Sections~\ref{sec:experiments} and~\ref{sec:nhanes} present the simulations
and NHANES analysis, and Section~\ref{sec:discussion} concludes.

\section{The Hilbert Mass-Addressed Parameterization}
\label{sec:preli}

HiMAP uses a fixed binary tree and conditional medians to separate common
probability labels from distribution-specific cells.
The Q representation retains physical locations along the tree, whereas the TL
representation records the same hierarchy in coordinates that remain valid
under affine aggregation.
Figures~\ref{fig:himap-address-construction} and~\ref{fig:himap-aggregation} introduce these two steps.

\subsection{Conditional-median mass addresses}
\label{subsec:overview-address}

Let $\mathcal P_r(\mathbb R^d)$ denote the Borel probability measures with finite $r$th moment, and let $W_r$ be the $r$-Wasserstein distance.
For a measurable map $T$, write $T_\#\mu$ for its pushforward.
In one dimension, a quantile index identifies a common amount of probability mass, while the quantile value gives its location for a particular distribution.
HiMAP separates these two roles in several dimensions by using a binary address $\sigma$ as the common mass label.

Write $[d]=\{1,\ldots,d\}$ and $\mathbb T_l=\{0,1\}^l$ for $l\geq0$, with $\mathbb T_0=\{\varnothing\}$.
The full address tree is $\mathbb T=\bigcup_{l\geq0}\mathbb T_l$.
We use $\sigma\in\mathbb T$ for a generic node, $\sigma_l\in\mathbb T_l$ for an address at level $l$, and $\sigma_L\in\mathbb T_L$ for a terminal address in a depth-$L$ truncation.
For $l\leq L$, the prefix of $\sigma_L$ at level $l$ is denoted by $\sigma_l=\sigma_L|_l$.
The two children of a generic node are $\sigma0$ and $\sigma1$.

Every node $\sigma$ has a split coordinate $j_\sigma\in[d]$ and an orientation $o_\sigma\in\{-1,1\}$ fixed before any distribution is observed.
The implemented block-cyclic Hilbert schedule visits every coordinate once in each complete block of $d$ levels.
Hence every $\sigma_L\in\mathbb T_L$ visits each coordinate at least $\lfloor L/d\rfloor$ times along its prefixes.
Within the initial incomplete block, level $l<d$ uses coordinate $l+1$.
A depth-$L$ tree with $L<d$ therefore splits only the first $L$ coordinates, although the empirical order retains all coordinates of every observation.

Starting from a distribution-specific root rectangle $K_\varnothing^\mu$, the encoder recursively associates a physical cell
$K_\sigma^\mu=\prod_{j=1}^d[a_{\sigma j}^\mu,b_{\sigma j}^\mu]$ with every node $\sigma$.
Within $K_\sigma^\mu$, the cut is the conditional median
\begin{equation}
q_{\mu,\sigma}
=\inf\left\{x\in\mathbb R:
\mu\bigl(\{\bu\in K_\sigma^\mu:u_{j_\sigma}\leq x\}\bigr)
\geq\frac12\mu(K_\sigma^\mu)\right\}.
\label{eq:conditional-median-address}
\end{equation}
The orientation $o_\sigma$ assigns the two geometric pieces to $K_{\sigma0}^\mu$ and $K_{\sigma1}^\mu$.
Under the encoder conditions below, the median is unique and the split face has zero probability.
Consequently, the two children receive equal mass.
After $L$ splits, every address $\sigma_L\in\mathbb T_L$ represents mass $2^{-L}$.
The address has the same probabilistic meaning for every distribution, although the corresponding physical cell varies with $\mu$.

For an empirical cloud, a node containing $N$ observations is divided by their ranks along the scheduled coordinate.
The two children receive $\lfloor N/2\rfloor$ and $\lceil N/2\rceil$ observations, with orientation determining their address order.
When the adjacent separating values are distinct, their midpoint provides a numerical cut that places no observation on the split face.
Repeating this construction yields terminal leaves containing one observation each.
For $n$ distinct observations, the maximum leaf depth is $\lceil\log_2 n\rceil$.
Reading the leaves by their binary addresses gives a complete order $U_{(1)},\ldots,U_{(n)}$ for every $n$.
Figure~\ref{fig:himap-address-construction} shows the construction for eight points in two dimensions.

\begin{figure}[t]
\centering
\begin{adjustbox}{max width=0.98\textwidth}
\begin{tikzpicture}[
  x=1cm,y=1cm,
  every node/.style={font=\scriptsize,text=black!88},
  paneltitle/.style={font=\footnotesize\bfseries,align=center},
  point/.style={circle,draw=black!75,fill=white,thick,inner sep=0pt,minimum size=4.2mm,font=\tiny},
  tree/.style={circle,draw=black!70,fill=white,thick,inner sep=0pt,minimum size=5mm,font=\tiny},
  leaf/.style={draw=black!65,rounded corners=1pt,fill=white,inner sep=2pt,align=center,font=\scriptsize},
  cutone/.style={draw=red!72!black,very thick},
  cuttwo/.style={draw=black!55,thick,dashed},
  cutthree/.style={draw=black!55,thick},
  path/.style={-{Stealth[length=1.8mm]},draw=blue!68!black,very thick},
  line cap=round,line join=round
]

\node[paneltitle] at (3.45,4.72) {(a) Conditional rank splits and Hilbert traversal};
\fill[black!3] (0.60,0.55) rectangle (3.58,2.11);
\fill[black!6] (0.60,2.11) rectangle (3.58,4.20);
\fill[black!3] (3.58,2.60) rectangle (6.30,4.20);
\fill[black!6] (3.58,0.55) rectangle (6.30,2.60);
\draw[black!70,thick] (0.60,0.55) rectangle (6.30,4.20);
\draw[cutone] (3.58,0.55)--(3.58,4.20);
\draw[cuttwo] (0.60,2.11)--(3.58,2.11);
\draw[cuttwo] (3.58,2.60)--(6.30,2.60);
\draw[cutthree] (0.60,1.00)--(3.58,1.00);
\draw[cutthree] (2.13,2.11)--(2.13,4.20);
\draw[cutthree] (5.24,2.60)--(5.24,4.20);
\draw[cutthree] (3.58,1.44)--(6.30,1.44);
\node[fill=white,inner sep=1pt,rotate=90] at (3.58,3.83)
  {$\widehat q_\varnothing$};

\draw[path] plot coordinates {
  (2.09,0.77) (2.09,1.55) (1.36,3.15) (2.85,3.15)
  (4.41,3.40) (5.77,3.40) (4.94,2.02) (4.94,1.00)
};

\node[point] at (0.80,1.31) {$1$};
\node[point] at (2.66,0.68) {$2$};
\node[point] at (1.33,3.62) {$3$};
\node[point] at (2.92,2.91) {$4$};
\node[point] at (4.25,1.13) {$5$};
\node[point] at (5.97,1.75) {$6$};
\node[point] at (4.38,3.44) {$7$};
\node[point] at (6.10,4.07) {$8$};

\node[anchor=south west,font=\tiny] at (0.64,0.58) {$000$};
\node[anchor=north east,font=\tiny] at (3.54,2.07) {$001$};
\node[anchor=north west,font=\tiny] at (0.64,4.16) {$010$};
\node[anchor=north west,font=\tiny] at (2.17,4.16) {$011$};
\node[anchor=north west,font=\tiny] at (3.62,4.16) {$100$};
\node[anchor=north west,font=\tiny] at (5.28,4.16) {$101$};
\node[anchor=north west,font=\tiny] at (3.62,2.56) {$110$};
\node[anchor=south west,font=\tiny] at (3.62,0.58) {$111$};
\node[font=\tiny,anchor=west] at (0.62,4.40)
  {root cut in red; later levels in dashed and solid black};

\node[paneltitle] at (11.85,4.72) {(b) Binary tree with singleton leaves};
\node[tree] (r) at (11.85,4.05) {$\varnothing$};
\node[tree] (a0) at (9.85,3.22) {$0$};
\node[tree] (a1) at (13.85,3.22) {$1$};
\node[tree] (a00) at (8.85,2.35) {$00$};
\node[tree] (a01) at (10.85,2.35) {$01$};
\node[tree] (a10) at (12.85,2.35) {$10$};
\node[tree] (a11) at (14.85,2.35) {$11$};
\draw[black!65,thick] (r)--(a0) (r)--(a1);
\draw[black!65,thick] (a0)--(a00) (a0)--(a01);
\draw[black!65,thick] (a1)--(a10) (a1)--(a11);

\node[leaf] (l000) at (8.35,0.95) {$000$\\$U_2$};
\node[leaf] (l001) at (9.35,0.95) {$001$\\$U_1$};
\node[leaf] (l010) at (10.35,0.95) {$010$\\$U_3$};
\node[leaf] (l011) at (11.35,0.95) {$011$\\$U_4$};
\node[leaf] (l100) at (12.35,0.95) {$100$\\$U_7$};
\node[leaf] (l101) at (13.35,0.95) {$101$\\$U_8$};
\node[leaf] (l110) at (14.35,0.95) {$110$\\$U_6$};
\node[leaf] (l111) at (15.35,0.95) {$111$\\$U_5$};
\draw[black!55,thick] (a00)--(l000) (a00)--(l001);
\draw[black!55,thick] (a01)--(l010) (a01)--(l011);
\draw[black!55,thick] (a10)--(l100) (a10)--(l101);
\draw[black!55,thick] (a11)--(l110) (a11)--(l111);
\node[align=center] at (11.85,0.16)
  {leaf order: $U_2,U_1,U_3,U_4,U_7,U_8,U_6,U_5$};

\end{tikzpicture}
\end{adjustbox}
\caption{Construction of an empirical mass address for $n=8$ points in two dimensions.
Panel (a) divides the ranks at every current node along its scheduled coordinate.
The recursion stops when each terminal cell contains one observation, and the blue path visits these cells in Hilbert address order.
Panel (b) shows the equivalent binary tree, whose leaves define the complete order $U_{(1)},\ldots,U_{(8)}$.
Numbers inside the points are their original input indices.
The binary addresses are common across distributions, whereas the physical cut locations depend on the input cloud.}
\label{fig:himap-address-construction}
\end{figure}

\subsection{Two representations on the common address}

The shared addresses support two different aggregation rules.
Q-HiMAP averages physical path locations with nonnegative weights, while
TL-HiMAP transforms relative split locations into unconstrained coordinates
that can be averaged with real weights summing to one and decoded to a
probability law.
Figure~\ref{fig:himap-aggregation} shows where the two branches separate.

\begin{figure}[t]
\centering
\begin{adjustbox}{max width=0.96\textwidth}
\begin{tikzpicture}[
  x=1cm,y=1cm,
  every node/.style={font=\scriptsize,text=black!88,align=center},
  block/.style={draw=black!68,thick,rounded corners=1.5pt,fill=white,minimum height=0.78cm,inner sep=4pt},
  qblock/.style={block,fill=blue!6},
  tlblock/.style={block,fill=red!5},
  arrow/.style={-{Stealth[length=1.8mm]},draw=black!68,thick},
  qarrow/.style={-{Stealth[length=1.8mm]},draw=blue!68!black,thick},
  tlarrow/.style={-{Stealth[length=1.8mm]},draw=red!68!black,thick},
  line cap=round,line join=round
]

\node[block,text width=2.15cm] (inputs) at (1.35,2.30)
  {input laws\\$\mu_1,\ldots,\mu_q$};
\node[block,text width=2.65cm] (address) at (4.55,2.30)
  {common mass address\\$\sigma\in\mathbb T$, mass $2^{-|\sigma|}$};
\draw[arrow] (inputs)--(address);

\node[qblock,text width=2.55cm] (qenc) at (8.20,3.35)
  {physical paths\\$Q_{\mu_1},\ldots,Q_{\mu_q}$};
\node[qblock,text width=2.85cm] (qavg) at (11.65,3.35)
  {positive path average\\$Q_\alpha=\sum_i\alpha_iQ_{\mu_i}$\\$\alpha_i\geq0$, $\sum_i\alpha_i=1$};
\node[qblock,text width=2.65cm] (qout) at (15.10,3.35)
  {Q-HiMAP barycenter\\$(Q_\alpha)_\#\operatorname{Unif}[0,1]$};

\node[tlblock,text width=2.55cm] (tlenc) at (8.20,1.25)
  {TL coordinates\\$\bg_i=\Phi(\mu_i)$};
\node[tlblock,text width=2.85cm] (tlavg) at (11.65,1.25)
  {affine average\\$\bg_\alpha=\sum_i\alpha_i\bg_i$\\$\alpha_i\in\mathbb R$, $\sum_i\alpha_i=1$};
\node[tlblock,text width=2.65cm] (tlout) at (15.10,1.25)
  {TL-HiMAP barycenter\\$D(\bg_\alpha)$};

\draw[qarrow] (address.east)--++(0.32,0)--++(0,1.05)--(qenc.west);
\draw[tlarrow] (address.east)--++(0.32,0)--++(0,-1.05)--(tlenc.west);
\draw[qarrow] (qenc)--(qavg);
\draw[qarrow] (qavg)--(qout);
\draw[tlarrow] (tlenc)--(tlavg);
\draw[tlarrow] (tlavg)--(tlout);

\node[font=\footnotesize\bfseries,text=blue!68!black,anchor=east]
  at (6.30,3.35) {spatial branch};
\node[font=\footnotesize\bfseries,text=red!68!black,anchor=east]
  at (6.30,1.25) {affine branch};

\end{tikzpicture}
\end{adjustbox}
\caption{The two HiMAP aggregation rules share the same mass addresses but average different objects.
Q-HiMAP averages physical path locations and is used with nonnegative weights.
TL-HiMAP averages Tree-logit coordinates and decodes every finite affine combination whose weights sum to one.}
\label{fig:himap-aggregation}
\end{figure}

Throughout the paper, \qhimap{} denotes the nonnegative spatial operation induced by $Q$.
The label \qaffine{} (empirical) is reserved for applying affine weights directly to common-rank empirical particles.
Q-affine is an empirical particle rule, whereas the population $d_{Q,2}$
barycenter uses nonnegative weights.

Let $\mathcal R=[-1/2,1/2]^d$ be the normalized root rectangle.
For $\bc,\bell\in\mathbb R^d$, define
$S_{\bc,\bell}(\bu)=\bc+\operatorname{diag}(e^{\bell})\bu$ and
$\mathcal M(\bc,\bell)=S_{\bc,\bell}(\mathcal R)$, where exponentiation is componentwise.
Suppose that $\mu$ has a nondegenerate compact rectangular support.
Its center $\bc_\mu$ and log side-length vector $\bell_\mu$ determine the normalized law
$\bar\mu=(S_{\bc_\mu,\bell_\mu}^{-1})_\#\mu$ on $\mathcal R$.

For the affine representation, fix a common buffer $\varepsilon\in(0,1/2)$ and define
\[
g_\varepsilon(a)
=\varepsilon+(1-2\varepsilon)(1+e^{-a})^{-1},
\qquad
\psi_\varepsilon(r)
=\log\frac{r-\varepsilon}{1-\varepsilon-r}.
\]
These maps are inverse bijections between $\mathbb R$ and $(\varepsilon,1-\varepsilon)$.
When the relative position of the conditional median lies in this interval, define
\[
r_{\mu,\sigma}
=\frac{q_{\mu,\sigma}-a_{\sigma,j_\sigma}^\mu}
{b_{\sigma,j_\sigma}^\mu-a_{\sigma,j_\sigma}^\mu},
\qquad
z_{\mu,\sigma}=\psi_\varepsilon(r_{\mu,\sigma}).
\]
Thus the TL encoder follows the order
$q_{\mu,\sigma}\mapsto r_{\mu,\sigma}\mapsto z_{\mu,\sigma}$ at every node.
Let $\bz_\mu=(z_{\mu,\sigma})_{\sigma\in\mathbb T}$ denote the complete coordinate sequence.

TL decoding recovers the relative split positions from the coordinates in the
reverse order.
Let $\bz=(z_\sigma)_{\sigma\in\mathbb T}$ be a real Tree-logit sequence.
The scalar $z_\sigma$ belongs to node $\sigma$ and specifies the relative position of the cut at that node.
Before the output law exists, use $\widetilde K_\sigma^{\bz}$ and $\widetilde q_\sigma^{\bz}$ for the normalized cell and cut reconstructed from $\bz$.
Set $\widetilde K_\varnothing^{\bz}=\mathcal R$.
The strict ancestors of $\sigma$ determine $\widetilde K_\sigma^{\bz}$, and $z_\sigma$ determines the cut inside this cell.
If
$\widetilde K_\sigma^{\bz}=\prod_{j=1}^d[a_{\sigma j}^{\bz},b_{\sigma j}^{\bz}]$, define
\[
r_\sigma^{\bz}=g_\varepsilon(z_\sigma),
\qquad
\widetilde q_\sigma^{\bz}
=a_{\sigma,j_\sigma}^{\bz}
+r_\sigma^{\bz}
\{b_{\sigma,j_\sigma}^{\bz}-a_{\sigma,j_\sigma}^{\bz}\}.
\]
After the orientation assigns the lower and upper geometric children to
address children $\sigma0$ and $\sigma1$, the tildes distinguish objects
constructed before the decoded law is identified.
Section~\ref{sec:representation} proves that these reconstructed objects are
the cells and conditional medians of the decoded law.

Let $B_1(t),B_2(t),\ldots$ be the fixed binary digits of $t\in[0,1)$, using one convention at dyadic ambiguities, and write
$A_l(t)=(B_1(t),\ldots,B_l(t))\in\mathbb T_l$ for its level-$l$ address.
For $\sigma_l\in\mathbb T_l$, the cylinder $I_{\sigma_l}=\{t:A_l(t)=\sigma_l\}$ has Lebesgue mass $2^{-l}$.

The cylinders form the distribution-free address system, whereas $\widetilde K_{\sigma_l}^{\bz}$ is the normalized geometry reconstructed from $\bz$ at address $\sigma_l$.

Figure~\ref{fig:recursive-address} makes this dictionary explicit.
It pairs the probability label and mass in $I_{\sigma_l}$ with the
distribution-specific physical region $K_{\sigma_l}^\mu$ and traces the cut
$q_{\mu,\sigma}$ through the relative coordinate $r_{\mu,\sigma}$ to the
unconstrained coordinate $z_{\mu,\sigma}$.

\begin{figure}[t]
\centering
\begin{adjustbox}{max width=\textwidth}
\begin{tikzpicture}[
  x=0.9cm,y=0.9cm,
  every node/.style={font=\footnotesize,text=black,align=center},
  address/.style={draw=blue!65!black,thick},
  geometry/.style={draw=orange!75!black,thick},
  cut/.style={draw=red!70!black,thick},
  highlight/.style={draw=blue!65!black,fill=blue!9,thick},
  ratio/.style={draw=teal!70!black,thick},
  arrow/.style={-{Stealth[length=2mm]},thick,draw=black!70},
  line cap=round,line join=round
]

\node[font=\footnotesize\bfseries] at (1.8,3.35)
  {(a) Binary mass address};
\draw[address] (0,1.35) rectangle (3.6,2.45);
\filldraw[highlight] (0.9,1.35) rectangle (1.8,2.45);
\foreach \x in {0.9,1.8,2.7} {\draw[address] (\x,1.35)--(\x,2.45);}
\node at (0.45,1.90) {$00$};
\node at (1.35,1.90) {$01$};
\node at (2.25,1.90) {$10$};
\node at (3.15,1.90) {$11$};
\node[font=\scriptsize] at (0,1.08) {$0$};
\node[font=\scriptsize] at (0.9,1.08) {$1/4$};
\node[font=\scriptsize] at (1.8,1.08) {$1/2$};
\node[font=\scriptsize] at (2.7,1.08) {$3/4$};
\node[font=\scriptsize] at (3.6,1.08) {$1$};
\node at (1.8,0.45)
  {$I_{01}=[1/4,1/2),\quad \operatorname{Leb}(I_{01})=2^{-2}$};

\draw[arrow] (3.85,1.90) -- node[above,font=\scriptsize,fill=white,inner sep=1pt]
  {$\sigma_2=01$} (4.75,1.90);

\node[font=\footnotesize\bfseries] at (7.15,3.35)
  {(b) Physical cells $K_{\sigma_2}^{\mu}$};
\filldraw[highlight] (5.0,1.72) rectangle (6.65,2.85);
\draw[geometry] (5.0,0.55) rectangle (9.3,2.85);
\draw[cut] (6.65,0.55)--(6.65,2.85);
\draw[cut] (5.0,1.72)--(6.65,1.72);
\draw[cut] (6.65,1.38)--(9.3,1.38);
\node at (5.82,1.15) {$K_{00}^{\mu}$};
\node at (5.82,2.36) {$K_{01}^{\mu}$};
\node[font=\scriptsize] at (5.82,2.05) {mass $2^{-2}$};
\node at (7.98,0.95) {$K_{10}^{\mu}$};
\node at (7.98,2.10) {$K_{11}^{\mu}$};
\node[rotate=90,font=\scriptsize,text=red!70!black,fill=white,inner sep=1pt]
  at (6.65,2.40) {$q_{\mu,\varnothing}$};

\draw[arrow] (9.55,1.90) -- node[above=2pt,pos=0.43,font=\scriptsize,fill=white,inner xsep=2pt,inner ysep=1pt]
  {encode split} (10.45,1.90);

\node[font=\footnotesize\bfseries] at (13.25,3.35)
  {(c) TL coordinate at node $\sigma$};
\draw[geometry] (10.75,2.05)--(15.75,2.05);
\draw[geometry] (10.75,1.83)--(10.75,2.27);
\draw[cut] (12.65,1.78)--(12.65,2.32);
\draw[geometry] (15.75,1.83)--(15.75,2.27);
\draw[ratio,<->] (10.75,1.45)--(12.65,1.45);
\node at (10.75,2.55) {$a_{\sigma,j_\sigma}^{\mu}$};
\node[text=red!70!black] at (12.65,2.55) {$q_{\mu,\sigma}$};
\node at (15.75,2.55) {$b_{\sigma,j_\sigma}^{\mu}$};
\node[text=teal!70!black] at (13.25,0.92)
  {$r_{\mu,\sigma}=\dfrac{q_{\mu,\sigma}-a_{\sigma,j_\sigma}^{\mu}}
  {b_{\sigma,j_\sigma}^{\mu}-a_{\sigma,j_\sigma}^{\mu}}$};
\node[text=teal!70!black] at (13.25,0.28)
  {$z_{\mu,\sigma}=\psi_\varepsilon(r_{\mu,\sigma})$};

\end{tikzpicture}
\end{adjustbox}
\caption{The mass-address dictionary at depth two.
Panel (a) selects the probability cylinder with address $01$ and mass $2^{-2}$.
Panel (b) attaches to that address a distribution-specific physical cell of the same mass.
Panel (c) converts a conditional-median cut into its relative position and Tree-logit coordinate.
Blue traces the common address, orange marks physical geometry, red marks median cuts, and teal marks the relative coordinate.}
\label{fig:recursive-address}
\end{figure}

The representation theorem in Section~\ref{sec:representation} shows that
the nested cells along each address have a singleton intersection.
Identifying that singleton with its point defines the normalized path
\[
\bar Q_{\bz}(t)=\bigcap_{l\geq0}\widetilde K_{A_l(t)}^{\bz},
\qquad
\bar D_\varepsilon(\bz)
=(\bar Q_{\bz})_\#\operatorname{Unif}[0,1].
\]
The safe buffer and balanced schedule provide the required uniform mesh contraction.

For $0<\theta<1$, define the weighted tree Hilbert space
\[
\mathcal H_\theta
=\left\{\bz:
\|\bz\|_{\mathcal H_\theta}^2
=\sum_{l=0}^\infty\theta^l2^{-l}
\sum_{\sigma_l\in\mathbb T_l}z_{\sigma_l}^2<\infty\right\}.
\]
By averaging squared coordinates within each level and discounting deeper
levels, these weights make the resulting separable Hilbert space a linear
domain for finite affine combinations.

\subsection{Coordinate space, decoder, and induced metrics}

The TL coordinate contains an explicit root block formed by the center $\bc$
and log width $\bell$.
Both quantities are recovered canonically from the support.
By contrast, the Q representation treats the composed physical path $Q_{\bg}$ as one spatial object rather than averaging $\bc$ and $\bell$ separately.

With fixed positive calibration constants $\eta_c$ and $\eta_\ell$, define
\[
\mathcal G
=\mathbb R^d\times\mathbb R^d\times\mathcal H_\theta,
\qquad
\|(\bc,\bell,\bz)\|_{\mathcal G}^2
=\eta_c\|\bc\|_2^2+\eta_\ell\|\bell\|_2^2
+\|\bz\|_{\mathcal H_\theta}^2.
\]
The structural tuple $\vartheta$ collects the dimension, schedule, $\varepsilon$, $\theta$, and metric calibration.
We keep this tuple fixed throughout the main paper, so we write
$\mathcal G$, $\mathcal X$, $D$, $\Phi$, and $d_{\rm TL}$ without a
structural subscript.
The supplement restores full subscripts when assumptions vary.
For $\bg=(\bc,\bell,\bz)\in\mathcal G$, set
\[
Q_{\bg}=S_{\bc,\bell}\circ\bar Q_{\bz},
\qquad
D(\bg)=(Q_{\bg})_\#\operatorname{Unif}[0,1],
\qquad
\mathcal X=D(\mathcal G).
\]
The physical decoder cell at address $\sigma$ is
\[
K_\sigma^{\bg}
:=S_{\bc,\bell}(\widetilde K_\sigma^{\bz}).
\]
Thus $\widetilde K_\sigma^{\bz}$ denotes normalized decoder geometry, while
$K_\sigma^{\bg}$ includes the root location and side lengths.
The representation theorem below shows that
$K_\sigma^{\Phi(\mu)}=K_\sigma^\mu$ for every encoded law.

The population encoder is initially defined on $\mathcal A^{\rm enc}$.
This class consists of laws with nondegenerate compact rectangular support whose normalized density $\bar f_\mu$ satisfies
$0<m_\mu\leq\bar f_\mu\leq M_\mu<\infty$ and $\varepsilon<m_\mu/(2M_\mu)$.
These bounds make every prescribed conditional median unique and keep its relative position inside the common buffer.

For $\mu\in\mathcal A^{\rm enc}$, the support rectangle gives $(\bc_\mu,\bell_\mu)$ and the recursive medians give $\bz_\mu$.
We collect them in
\[
\Phi(\mu)=(\bc_\mu,\bell_\mu,\bz_\mu)
\qquad\text{and set}\qquad
Q_\mu=Q_{\Phi(\mu)}.
\]
Applying the same coordinatewise normalization to every encoder cell and its
split coordinate gives
\[
K_\sigma^{\bar\mu}
=S_{\bc_\mu,\bell_\mu}^{-1}(K_\sigma^\mu),
\qquad
q_{\bar\mu,\sigma}
=e^{-\ell_{\mu,j_\sigma}}
\{q_{\mu,\sigma}-c_{\mu,j_\sigma}\}.
\]
Since the normalization is increasing in each coordinate, the normalized cell
and cut are exactly those obtained by applying the encoder directly to
$\bar\mu$.
Thus the trees of $\mu$ and $\bar\mu$ correspond node by node under $S_{\bc_\mu,\bell_\mu}^{-1}$.

The decoder starts instead from $\bz_\mu$ and reconstructs provisional cells and cuts.
The recursive definitions give
\[
\widetilde K_\sigma^{\bz_\mu}=K_\sigma^{\bar\mu},
\qquad
\widetilde q_\sigma^{\bz_\mu}=q_{\bar\mu,\sigma},
\qquad \sigma\in\mathbb T.
\]
Section~\ref{sec:representation} then proves that the reconstructed path has law $\bar\mu$ and that every decoded law recovers its generating coordinate.
We then use $\Phi=D^{-1}$ and $Q_\mu=Q_{\Phi(\mu)}$ throughout $\mathcal X$.

After this extension, the common tree supports two distances on $\mathcal X$.
For $r\geq1$, define
\begin{align*}
d_{Q,r}(\mu,\nu)
&=\|Q_\mu-Q_\nu\|_{L^r([0,1];\mathbb R^d)},
\\
d_{\rm TL}(\mu,\nu)
&=\|\Phi(\mu)-\Phi(\nu)\|_{\mathcal G}.
\end{align*}
The Q distance compares physical locations at a shared mass address, whereas
the TL distance compares root and relative split coordinates.
The two distances define distinct barycentric objectives even when all weights
are nonnegative.

This distinction also persists empirically.
For observations $U_1,\ldots,U_n\in\mathbb R^d$, recursive rank halving
produces a permutation $\pi(U_{1:n})$.
The permutation uses only the fixed schedule and comparisons among
observations.
Empirical Q-HiMAP averages the ordered observations directly.
Using the same ranks, TL-HiMAP estimates the root and finite-depth split
coordinates that define the population TL barycenter.
Section~\ref{sec:estimation} introduces a separate within-cell refinement that converts a truncated TL law into an equal-weight particle output without changing the population geometry.

The constants $\eta_c$ and $\eta_\ell$ calibrate the root terms of $d_{\rm TL}$, while $\theta$ determines the tree space and its depth weighting.
These choices affect the metric and its stability constants but not the affine
coordinate average, because every summand is measured in the same squared
Hilbert norm.

\section{Barycenters and Coordinate Geometry}
\label{sec:representation}

Section~\ref{sec:preli} defined a common mass address and its Q and TL
representations.
The population theory below covers barycenters for both representations,
transport stability and model coverage, and TL Fr\'echet regression.
Component results and complete proofs appear in the supplement.

\subsection[Mass-addressed representation and inverse]
{Mass-addressed representation and inverse}

Put $\rho=(1-\varepsilon)^{1/d}$ for the implemented block-cyclic schedule.
A general pathwise-balanced schedule changes only the contraction constants.

\begin{theorem}[Mass-addressed representation]
\label{thm:main-representation}
Fix $\varepsilon\in(0,1/2)$ and the block-cyclic schedule.
For every $\bg\in\mathcal G$, nested address cells contract uniformly at rate
$O(\rho^L)$ and define a measurable path map.
Every depth-$L$ address cell has probability $2^{-L}$ under $D(\bg)$.
For every $\mu\in\mathcal A^{\rm enc}$,
$D\{\Phi(\mu)\}=\mu$.
Conversely, the canonical root and all conditional-median splits of a decoded
law are uniquely recoverable and
\[
\Phi\{D(\bg)\}=\bg.
\]
Thus $D$ is a bijection from $\mathcal G$ onto
$\mathcal X$, and
$(\mathcal X,d_{\rm TL})$ is complete and separable.
\end{theorem}

Theorem~\ref{thm:main-representation} identifies the Hilbert coordinate with
the distribution it decodes.
The buffer and block-cyclic schedule are common structural choices for the
model class and do not restrict the affine weights used later.
Every address has the same probability meaning across laws, and every decoded
output can be re-encoded without a reference distribution.
Appendix~\ref{app:proof-main-representation} gives the complete proof.

\begin{corollary}[Univariate compatibility]
\label{cor:main-univariate}
In dimension one, $Q_\mu(t)=F_\mu^{-1}(t)$ almost everywhere.
Consequently, $d_{Q,r}=W_r$, and the positive Q barycenter is the usual
one-dimensional Wasserstein barycenter.
For $\sigma_l\in\mathbb T_l$, let $k(\sigma_l)$ be the integer represented by its binary digits and set
\[
p^-_{\sigma_l}=\frac{k(\sigma_l)}{2^l},
\qquad
p^0_{\sigma_l}=\frac{2k(\sigma_l)+1}{2^{l+1}},
\qquad
p^+_{\sigma_l}=\frac{k(\sigma_l)+1}{2^l}.
\]
With the support endpoints used at probability levels zero and one, the corresponding TL coordinate is
\[
z_{\mu,\sigma_l}
=\psi_\varepsilon\left\{
\frac{F_\mu^{-1}(p^0_{\sigma_l})-F_\mu^{-1}(p^-_{\sigma_l})}
{F_\mu^{-1}(p^+_{\sigma_l})-F_\mu^{-1}(p^-_{\sigma_l})}
\right\}.
\]
\end{corollary}

Thus the Q representation reduces to the classical quantile function, whereas
the TL representation reduces to logit-transformed relative locations of
nested dyadic quantiles.
The TL coordinates form a deterministic quantile-pyramid coordinate system
\citep{HjortWalker2009QuantilePyramids,AnMacEachern2024DependentQuantilePyramids}.
They may be averaged with real weights that sum to one and then decoded, but
the induced TL metric is generally different from $W_r$.
Proposition~\ref{prop:univariate-reduction} in the supplement states the full reduction, and Appendix~\ref{app:proof-univariate-reduction} gives its proof.

\subsection[Two geometries, closure, and barycenters]
{Two geometries, closure, and barycenters}
\label{sec:barycenter-frechet}

The Q representation preserves convex combinations of physical paths, whereas
the TL representation preserves affine combinations through its complete real
Hilbert coordinate space.
The Q and TL representations admit different weights and therefore define
different least-squares barycenters.

\begin{theorem}[Two closure laws and induced barycenters]
\label{thm:main-barycenters}
Let $\mu_1,\ldots,\mu_q\in\mathcal X$ and suppose
$\Lambda=\sum_i\lambda_i>0$.
Put $\alpha_i=\lambda_i/\Lambda$.
If every $\lambda_i\geq0$, the Q image is convex and the unique minimizer of
$\sum_i\lambda_i d_{Q,2}^2(\mu_i,\nu)$ satisfies
\[
Q_{\mu_Q^\oplus}=\sum_i\alpha_iQ_{\mu_i}.
\]
For arbitrary real $\lambda_i$, the TL image is affine and the unique
minimizer of $\sum_i\lambda_i d_{\rm TL}^2(\mu_i,\nu)$ is
\[
\mu_{\rm TL}^\oplus
=D\left\{\sum_i\alpha_i\Phi(\mu_i)\right\}.
\]
\end{theorem}

The two outputs are generally different even for positive weights.
Q-HiMAP averages physical widths arithmetically, whereas TL-HiMAP averages
log widths linearly.
The resulting Q formula defines a spatial positive barycenter, while the TL
formula defines an affine barycenter under the induced TL metric.
Even with signed coefficients, the TL output remains a probability law that
can be encoded again in the same coordinate system.
Each formula is the exact minimizer for its displayed induced-metric
objective, while Section~\ref{sec:experiments} evaluates both outputs under
the standard $W_2$ criterion.
Appendix~\ref{app:proof-main-barycenters} gives the complete proof.

\begin{corollary}[Exact affine compositionality]
\label{cor:main-compositionality}
Partition the inputs into groups $G_k$.
Let $\sum_{i\in G_k}a_{ki}=1$ and $\sum_kb_k=1$, and define
$\nu_k=D\{\sum_{i\in G_k}a_{ki}\Phi(\mu_i)\}$.
Then
\[
D\left\{\sum_k b_k\Phi(\nu_k)\right\}
=D\left\{\sum_k\sum_{i\in G_k}b_ka_{ki}
\Phi(\mu_i)\right\}.
\]
\end{corollary}

An intermediate TL output therefore retains the coordinates required by the
next affine operation.
The same result supports distributed and hierarchical aggregation without
changing the final population output.
Appendix~\ref{app:proof-main-compositionality} gives the complete proof.

\subsection[Transport stability and model coverage]
{Transport stability and model coverage}

The common mass address links both induced metrics to Wasserstein distance.
Continuity of the TL encoder in the reverse direction requires a regular class
of distributions.
We call a class uniformly regular at fixed depths when, at each fixed depth,
its root widths and densities have common positive lower and finite upper
bounds, its split ratios stay uniformly inside the buffer, and its conditional
medians have a common local identification modulus.
These conditions prevent degenerate recursive cells and make the split
locations stable under small probability perturbations.
Assumption~\ref{ass:uniform-fixed-depth} gives the precise constants.
Write $\|\mu-\nu\|_{\rm TV}=\sup_A|\mu(A)-\nu(A)|$.

\begin{theorem}[Transport comparison and encoder--decoder continuity]
\label{thm:main-stability}
The following statements hold.

\emph{(a) Spatial transport comparison.}
For $r\geq1$, $d_{Q,r}$ is a metric on $\mathcal X$ and
\begin{equation}
W_r(\mu,\nu)\leq d_{Q,r}(\mu,\nu).
\label{eq:main-q-upper-bound}
\end{equation}

\emph{(b) Forward TL continuity.}
If $\rho^2<\theta<1$ and the log widths of $\bg$ and $\bg'$ are bounded above by
$B$, then
\begin{equation}
W_2\{D(\bg),D(\bg')\}
\leq C_B\|\bg-\bg'\|_{\mathcal G},
\label{eq:main-decoder-stability}
\end{equation}
where $C_B$ depends only on the structural parameters, calibration, and $B$.

\emph{(c) Inverse continuity on regular classes.}
Let $\mathcal C\subset\mathcal A^{\rm enc}$ be uniformly regular at fixed
depths.
Let $P_L$ retain the root and split levels $0,\ldots,L-1$, and put
\[
\tau_L=\sup_{\mu\in\mathcal C}
\|(I-P_L)\bz_\mu\|_{\mathcal H_\theta}.
\]
For every fixed $L$, there are constants $A_L<\infty$ and $v_L>0$ such that
\begin{equation}
\|P_L\Phi(\mu)-P_L\Phi(\nu)\|_{\mathcal G}
\leq A_L\|\mu-\nu\|_{\rm TV},
\qquad
d_{\rm TL}(\mu,\nu)
\leq A_L\|\mu-\nu\|_{\rm TV}+2\tau_L,
\label{eq:main-inverse-continuity}
\end{equation}
whenever $\mu,\nu\in\mathcal C$ and
$\|\mu-\nu\|_{\rm TV}\leq v_L$.
If $\tau_L\to0$, then
\[
\lim_{a\downarrow0}\sup_{\substack{\mu,\nu\in\mathcal C\\
\|\mu-\nu\|_{\rm TV}\leq a}}
d_{\rm TL}(\mu,\nu)=0.
\]
\end{theorem}

The final conclusion of Theorem~\ref{thm:main-stability}(c) requires
$\tau_L\to0$ uniformly over $\mathcal C$.
Membership $\bz_\mu\in\mathcal H_\theta$ gives only pointwise tail convergence
for each fixed $\mu$, not uniform convergence over the class.
The following common normalized density bounds provide a sufficient uniform
condition.
Suppose $0<m\leq\bar f_\mu\leq M<\infty$ for every $\mu\in\mathcal C$ and
$\varepsilon<m/(2M)$.
Then every split ratio lies in $[r_\star,1-r_\star]$, where
$r_\star=m/(2M)$.
Set
$K_{m,M,\varepsilon}
=\max_{r\in[r_\star,1-r_\star]}|\psi_\varepsilon(r)|$.
Then $|z_{\mu,\sigma}|\leq K_{m,M,\varepsilon}$ uniformly in $\mu$ and
$\sigma$, and
\[
\tau_L\leq
K_{m,M,\varepsilon}\left\{\frac{\theta^L}{1-\theta}\right\}^{1/2}
\longrightarrow0.
\]
Consequently, every uniformly regular class satisfying these depth-independent
normalized density bounds obeys
\[
\lim_{\delta\downarrow0}
\sup_{\substack{\mu,\nu\in\mathcal C\\
\|\mu-\nu\|_{\rm TV}\leq\delta}}
d_{\rm TL}(\mu,\nu)=0.
\]
Classes whose TL coordinates vanish after a common finite depth give another
simple example.
The supplement derives the bound and gives the more general compactness
interpretation.

In dimension one, part (a) is an equality by
Corollary~\ref{cor:main-univariate}.
In several dimensions, $d_{Q,r}$, $d_{\rm TL}$, and Wasserstein distance are
distinct.
Parts (b) and (c) connect TL coordinates and probability laws in both
directions on uniformly regular population classes.
Section~\ref{sec:estimation} gives a separate fixed-depth analysis for
empirical atomic laws.
Appendix~\ref{app:proof-main-stability} gives the complete proof.

For model coverage, allow the buffer to vary and write
$\vartheta(\varepsilon)$.
Given $\mu\in\mathcal P_2(\mathbb R^d)$, project its tail onto a bounded ball,
smooth the result by cube-uniform convolution, and add a uniform floor on the
containing rectangle.
The resulting density is bounded and positive on its canonical root.
Let $\mu^{[A]}$ denote radial projection onto the radius-$A$ ball and set
$\mathcal B_{A,b}=[-A-b,A+b]^d$.

\begin{theorem}[Wasserstein approximation by buffered HiMAP models]
\label{thm:main-model-coverage}
\mbox{}\par\noindent
Let $\mu\in\mathcal P_2(\mathbb R^d)$ and fix $A,b>0$.
Choose $\delta\in(0,b)$ and $\omega\in(0,1)$, and define
\[
\nu_{A,b,\delta,\omega}
=(1-\omega)\{\mu^{[A]}*\operatorname{Unif}[-\delta,\delta]^d\}
+\omega\operatorname{Unif}(\mathcal B_{A,b}).
\]
For some $\varepsilon_\star\in(0,1/2)$, depending on
$(A,b,\delta,\omega)$, this law belongs to
$\mathcal A_{\vartheta(\varepsilon_\star)}^{\rm enc}$, and
\begin{equation}
W_2(\mu,\nu_{A,b,\delta,\omega})
\leq
\left\{\int (\|\bx\|_2-A)_+^2\,d\mu(\bx)\right\}^{1/2}
+\delta\sqrt{d/3}+2\sqrt d\,(A+b)\sqrt\omega .
\label{eq:model-coverage-bound}
\end{equation}
Consequently,
\begin{equation}
\overline{\bigcup_{0<\varepsilon<1/2}
\mathcal X_{\vartheta(\varepsilon)}}^{\,W_2}
=\mathcal P_2(\mathbb R^d).
\label{eq:model-coverage-density}
\end{equation}
\end{theorem}

The approximation applies to finite atomic laws, lower-dimensional supports,
and unbounded distributions with a finite second moment.
The union in \eqref{eq:model-coverage-density} lets the buffer adapt to the
regularization level.
Decoder stability and estimation continue to use one fixed buffer, so their
constants are interpreted within a fixed model class.
Appendix~\ref{app:proof-main-model-coverage} gives the complete proof.

\subsection[Fr\'echet regression in TL coordinates]
{Fr\'echet regression in TL coordinates}

Let $(X,Y)$ be a random pair with $X\in\mathbb R^p$ and let $Y$ be measurable
for the Borel sigma-field generated by $d_{\rm TL}$ on $\mathcal X$.
Then $G=\Phi(Y)$ is a strongly measurable random element of the separable
Hilbert space $\mathcal G$.
Under the TL metric, the conditional Fr\'echet target is
\begin{align*}
m_\oplus(\bx)
&\in\arg\min_{\nu\in\mathcal X}
E\{d_{\rm TL}^2(Y,\nu)\mid X=\bx\},\\
M_w(\nu,\bx)
&=E\{w(X,\bx)d_{\rm TL}^2(Y,\nu)\},
\qquad E\{w(X,\bx)\}=1.
\end{align*}
Global and local Fr\'echet regression use the second objective as a weighted
surrogate for the first.
The global weight comes from linear prediction around the predictor mean,
whereas the local weight satisfies the moment equations of a local linear fit.
Let $\boldsymbol\mu_X=E(X)$ and suppose
$\Sigma_X=\operatorname{Var}(X)$ is nonsingular.
For a kernel $K$ and bandwidth $h>0$, define
$K_h(\bv)=h^{-p}K(\bv/h)$,
$R_h(\bu;\bx)=(1,(\bu-\bx)^\top/h)^\top\in\mathbb R^{p+1}$, and
$M_h(\bx)=E\{K_h(X-\bx)R_h(X;\bx)R_h(X;\bx)^\top\}$.
When $M_h(\bx)$ is nonsingular, the two weights are
\begin{align*}
w_G(X,\bx)
&=1+(X-\boldsymbol\mu_X)^\top\Sigma_X^{-1}(\bx-\boldsymbol\mu_X),
\\
w_{{\rm loc},h}(X,\bx)
&=\boldsymbol e_1^\top M_h(\bx)^{-1}R_h(X;\bx)K_h(X-\bx),
\end{align*}
where $\boldsymbol e_1=(1,0,\ldots,0)^\top\in\mathbb R^{p+1}$.
Each weight has expectation one and may be negative under global extrapolation
or local linear boundary
correction.
The regression problem therefore requires affine closure
\citep{PetersenMuller2019FrechetRegression}.

For a measurable normalized weight $w(X,\bx)$, assume
\[
E|w(X,\bx)|<\infty,
\qquad
E\{|w(X,\bx)|\|G\|_{\mathcal G}^2\}<\infty.
\]

\begin{corollary}[Population TL Fr\'echet regression]
\label{thm:main-population-regression}
The weighted objective above has the unique minimizer
\[
m_{\rm TL}(\bx)=D\bigl[E\{w(X,\bx)G\}\bigr].
\]
In particular, suppose the conditional law is defined at $\bx$ and
$E(\|G\|_{\mathcal G}^2\mid X=\bx)<\infty$.
Then the conditional target is unique and equals $D[E(G\mid X=\bx)]$.
\end{corollary}

Corollary~\ref{thm:main-population-regression} therefore reduces each
prediction to one Hilbert-valued average and one decode.
For local regression, the finite bandwidth coordinate target is
$\bg_{{\rm loc},h}(\bx)=E\{w_{{\rm loc},h}(X,\bx)G\}$.
The difference between this finite-bandwidth coordinate and
$E(G\mid X=\bx)$ is the local linear smoothing bias.
Appendix~\ref{app:proof-main-population-regression} gives the complete proof
and the corresponding population formulas.
The supplement extends this result to estimated response laws:
Propositions~\ref{thm:regression-transfer} and
\ref{thm:joint-response-regime} give error transfer under joint sampling,
Corollary~\ref{cor:practical-residual-regression} controls the particle output,
and Proposition~\ref{prop:oracle-rates} gives the global $m^{-1/2}$ and local
$h^2+(mh^p)^{-1/2}$ oracle rates.
Section~\ref{sec:estimation} replaces these population coordinates and
expectations with finite point-cloud encoders and empirical affine weights.

\section{Estimation and Finite-Atom Computation}
\label{sec:estimation}

Section~\ref{sec:representation} describes the population paths and coordinates.
The empirical Q and TL procedures first remove the same near-degenerate
coordinates from every input cloud and then use the same complete recursive
median order.
Q-HiMAP averages observations at common ordered indices without estimating a
root or numerical split locations.
TL-HiMAP uses the order to estimate a finite tree whose zero-tail continuous
completion approximates the population law.
When an application requires exactly $n$ equally weighted output particles, a within-cell refinement places one particle at each ordered position.
Sections~\ref{app:detail-estimation}--\ref{app:statistical-proofs} give the
fixed-depth assumptions and complete proofs.

\subsection[Two empirical procedures]
{Two empirical procedures}
\label{subsec:atomic-output}

The coordinate screen is common to the whole aggregation problem.
For coordinate $j$, let $R_{ij}$ be its sample range in cloud $i$ and
$R_{\cdot j}=\max_iR_{ij}$.
With the default tolerance $\tau_{\rm dim}=10^{-8}$, retain
$\mathcal J_{\tau}=\{j:\min_iR_{ij}>\tau_{\rm dim}\max(1,R_{\cdot j})\}$ and
project every cloud onto these coordinates before constructing its order.
If no coordinate remains, return the method-admissible weighted point law at
the empirical centers.  We reuse $d$ for the active dimension.  The results
below apply in this common active space; the set is eventually constant when
the population range ratios are separated from the threshold.  No reported
experiment removes a coordinate.

For cloud $i$, let $\pi_i=\pi(U_{i,1:n})$ be the permutation obtained by
recursively halving the ranks until singleton leaves, and write
$U_{i,(k)}=U_{i,\pi_i(k)}$.
The complete order is defined for every $n$ and uses only the fixed schedule
and pairwise comparisons among observations.
Constructing this order requires neither a root box nor numerical split values.
The supplement specifies the deterministic cyclic lexicographic rule used
when an active coordinate is tied.
For nonnegative weights $\alpha_i$ summing to one, the Q procedure stops after
this common-order step and returns
\begin{equation}
\widetilde\mu_{Q,n}
=\frac1n\sum_{k=1}^n
\delta_{\sum_{i=1}^q\alpha_iU_{i,(k)}}.
\label{eq:main-practical-q-output}
\end{equation}

If the input clouds have unequal sizes, they may first be represented on a
common deterministic equal mass grid of size $n$.
Proposition~\ref{prop:unequal-grid} defines this preprocessing step and bounds
the discrepancy from each input empirical law.

The TL procedure next chooses $L$ with $2^L\leq n$.
For each cloud it takes the coordinatewise sample range as the empirical root.
At a node containing $N$ observations, it assigns the first
$\lfloor N/2\rfloor$ ordered observations to address child zero and the
remainder to child one.
The physical cut lies midway between the two adjacent separating order
statistics.
If $\widehat r_\sigma^{\rm cand}$ is the resulting relative split, define
\[
\widehat r_\sigma
=\Pi_\varepsilon(\widehat r_\sigma^{\rm cand}),
\qquad
\widehat z_\sigma=\psi_\varepsilon(\widehat r_\sigma).
\]
The finite clipping map $\Pi_\varepsilon$, specified in the supplement, maps
every candidate into
$(\varepsilon,1-\varepsilon)$, where the safe logit is finite.
Put $\mathbb T_{<L}=\bigcup_{l=0}^{L-1}\mathbb T_l$ and define the finite
coordinate space
\[
\mathcal G_L
=\mathbb R^d\times\mathbb R^d\times\mathbb R^{\mathbb T_{<L}}.
\]
Let $\widehat{\bg}_{i,n,L}\in\mathcal G_L$ contain the root center, root log
widths, and split logits through depth $L-1$.
For $\bg_L=(\bc,\bell,(z_\sigma)_{\sigma\in\mathbb T_{<L}})\in\mathcal G_L$,
define its zero-tail completion by
\[
\bg^{(L)}:=\iota_L(\bg_L)
=(\bc,\bell,\bz^{(L)})\in\mathcal G,
\qquad
z_\sigma^{(L)}=
\begin{cases}
z_\sigma,&|\sigma|<L,\\
0,&|\sigma|\geq L.
\end{cases}
\]
The continuous completion of the finite tree is the probability law
\begin{equation}
D_L^{\rm cont}(\bg_L)
:=D(\bg^{(L)})
=2^{-L}\sum_{\sigma\in\mathbb T_L}
\operatorname{Unif}(K_\sigma^{\bg^{(L)}}),
\label{eq:main-continuous-finite-decoder}
\end{equation}.
If $\bg_L=(\bc,\bell,\bz_{<L})$, then
$K_\sigma^{\bg^{(L)}}
=S_{\bc,\bell}(\widetilde K_\sigma^{\bz^{(L)}})$ is the physical depth-$L$
terminal rectangle.
The second equality follows because every completed split logit is zero, so
all later relative split ratios equal $g_\varepsilon(0)=1/2$.
Thus $D_L^{\rm cont}$ assigns mass $2^{-L}$ uniformly within each terminal
rectangle.
The map $\iota_L$ is linear, so it commutes with every affine coordinate
average.

The complete order assigns every position $k$ to a depth-$L$ terminal address
$\lambda_{n,L}(k)$.
Let $\widehat{\boldsymbol L}_{i,k}$ and
$\widehat{\boldsymbol U}_{i,k}$ be the lower and upper endpoint vectors of
this cell.
With coordinatewise operations, define
\[
\boldsymbol s_{i,k}^{\rm raw}
=
\operatorname{diag}(\widehat{\boldsymbol U}_{i,k}-\widehat{\boldsymbol L}_{i,k})^{-1}
\{U_{i,(k)}-\widehat{\boldsymbol L}_{i,k}\},
\qquad
\widetilde{\boldsymbol s}_{i,k}
=\Pi_\varepsilon^{(d)}(\boldsymbol s_{i,k}^{\rm raw}),
\qquad
\bh_{i,k}=\psi_\varepsilon^{(d)}(\widetilde{\boldsymbol s}_{i,k}).
\]
This inward clip makes every within-cell logit finite, including at a
sample-range endpoint.
Collect these rows as the within-cell refinement array
$H_i=(\bh_{i,1}^{\mathsf T},\ldots,\bh_{i,n}^{\mathsf T})^{\mathsf T}
=\widehat{\bh}_{i,n,L}\in\mathbb R^{n\times d}$.
For $\bg_L\in\mathcal G_L$, let
$\boldsymbol L_{\lambda_{n,L}(k)}^{\bg^{(L)}}$ and
$\boldsymbol U_{\lambda_{n,L}(k)}^{\bg^{(L)}}$ be the endpoint vectors of the
depth-$L$ cell reconstructed from $\bg_L$ at address $\lambda_{n,L}(k)$.
For $\bh\in\mathbb R^{n\times d}$ with $k$th row $\bh_k$, define
\begin{equation}
\bx_k(\bg_L,\bh)
=\boldsymbol L_{\lambda_{n,L}(k)}^{\bg^{(L)}}
+\operatorname{diag}\!\left\{
\boldsymbol U_{\lambda_{n,L}(k)}^{\bg^{(L)}}
-\boldsymbol L_{\lambda_{n,L}(k)}^{\bg^{(L)}}
\right\}g_\varepsilon^{(d)}(\bh_k).
\label{eq:main-residual-atom-decoder}
\end{equation}
Thus $\bg_L$ determines the terminal cell, while $g_\varepsilon^{(d)}(\bh_k)$
gives the relative position of particle $k$ inside that cell.
The empirical decoder returns
\begin{equation}
\widetilde D_{n,L}(\bg_L,\bh)
=\frac{1}{n}\sum_{k=1}^n\delta_{\bx_k(\bg_L,\bh)}.
\label{eq:main-residual-decoder}
\end{equation}
Thus the TL output under arbitrary affine weights summing to one is
\begin{equation}
\widetilde\mu_{{\rm TL},n,L}(\alpha)
=\widetilde D_{n,L}\left(
\sum_{i=1}^q\alpha_i\widehat{\bg}_{i,n,L},
\sum_{i=1}^q\alpha_iH_i\right).
\label{eq:main-practical-tl-output}
\end{equation}
The continuous TL law with the same finite tree coordinate is
\begin{equation}
\widehat\mu_{{\rm TL},n,L}^{\rm cont}(\alpha)
=D_L^{\rm cont}\left(\sum_{i=1}^q
\alpha_i\widehat{\bg}_{i,n,L}\right).
\label{eq:main-continuous-tl-output}
\end{equation}
The practical Q and TL procedures both return $n$ equally weighted particles,
counted with multiplicity, but they aggregate different empirical objects.
The array $H_i$ serves only to reconstruct each output particle inside its
decoded terminal cell and is not part of the population TL coordinate or the
metric $d_{\rm TL}$.

\begin{algorithm}[H]

\caption{Empirical HiMAP barycenter computation.}
\label{alg:empirical-himap}
\begin{algorithmic}[1]
\Require Clouds $U_{i,1:n}\subset\mathbb R^d$, $i=1,\ldots,q$, weights
$\alpha_i$ with $\sum_i\alpha_i=1$,
method $M\in\{Q,{\rm TL}\}$, and, for $M={\rm TL}$, depth $L$ and buffer
$\varepsilon$
\State apply the common coordinate screen $\mathcal J_\tau$; if it is empty,
enforce the method's weight restriction and return the weighted point law at
the empirical centers
\For{each input cloud $i$}
  \State compute the complete recursive order $\pi_i$ and
  $U_{i,(k)}=U_{i,\pi_i(k)}$
\EndFor
\If{$M=Q$}
  \State require $\alpha_i\geq0$ and return the common-rank average in
  \eqref{eq:main-practical-q-output}
\Else
  \For{each input cloud $i$}
    \State construct its sample-range root, depth-$L$ split logits
    $\widehat{\bg}_{i,n,L}$, and rank-indexed within-cell coordinates $H_i$
  \EndFor
  \State set $\widehat{\bg}=\sum_i\alpha_i\widehat{\bg}_{i,n,L}$ and
  $\widehat{\bh}=\sum_i\alpha_iH_i$
  \State return $\widetilde D_{n,L}(\widehat{\bg},\widehat{\bh})$ as in
  \eqref{eq:main-practical-tl-output}
\EndIf
\end{algorithmic}
\end{algorithm}

Algorithm~\ref{alg:empirical-himap} first computes the common ordering and then
applies the method-specific Q or TL step.
The complete recursive order ranks all $n$ observations independently of the
chosen TL depth.
If $n=2^L$, each depth-$L$ cell receives one ordered position.
For general $n$, integer halving assigns either $\lfloor n/2^L\rfloor$ or
$\lceil n/2^L\rceil$ positions to each cell, and the within-cell rows distinguish
their locations within that cell.
In several dimensions, split values alone do not recover the point in a
singleton cell.
The finite particle decoder therefore retains $H_i$ at one fixed depth.
Once the input distributions are encoded, their TL coordinates can be reused
for different barycentric weights.

\FloatBarrier
\subsection[Error decompositions for the two empirical barycenters]
{Error decompositions for the two empirical barycenters}
\label{sec:finite-barycenters}

The computed Q estimator uses neither $\widehat{\bg}_{i,n,L}$ nor $H_i$ and
returns an atomic law, whereas the population Q barycenter is formulated on
$\mathcal X$.
The proof connects the empirical and population objects through a continuous
law built from the same recursive order.
For each input cloud, form the auxiliary continuous tree law
$\widehat\mu_{i,n,L}^{\rm cont}$.
All split logits at levels $L,L+1,\ldots$ are set to zero, so the decoder
bisects the remaining intervals at their midpoints.
Consequently,
\[
\widehat\mu_{i,n,L}^{\rm cont}
=D_L^{\rm cont}(\widehat{\bg}_{i,n,L})
=2^{-L}\sum_{\sigma\in\mathbb T_L}
\operatorname{Unif}\{K_\sigma^{\iota_L(\widehat{\bg}_{i,n,L})}\}.
\]
Depending only on the fitted tree coordinate
$\widehat{\bg}_{i,n,L}$ rather than $H_i$, the auxiliary law replaces the atoms of the empirical law
$n^{-1}\sum_{k=1}^n\delta_{U_{i,k}}$ by uniform mass within the fitted
terminal rectangles and assigns every rectangle mass $2^{-L}$.
It also differs from the unknown law $\mu_i$ because its root and splits are
estimated from the sample and its density is constant within each terminal
rectangle, a discrepancy measured by
$d_{Q,2}(\widehat\mu_{i,n,L}^{\rm cont},\mu_i)$ in
\eqref{eq:main-q-finite-error}.
The corresponding reconstruction of ordered particle $U_{i,(k)}$ is
$U_{i,(k)}^{\rm rec}=\bx_k(\widehat{\bg}_{i,n,L},H_i)$.
Define the root-mean-square error of this particle reconstruction by
\[
\{e_{i,n,L}^{\rm rec}\}^2=\frac1n\sum_{k=1}^n
\|U_{i,(k)}-U_{i,(k)}^{\rm rec}\|_2^2.
\]
The same fitted terminal cell is used to encode and decode the particle.
Since $g_\varepsilon^{(d)}(\bh_{i,k})
=\widetilde{\boldsymbol s}_{i,k}$,
\[
U_{i,(k)}^{\rm rec}-U_{i,(k)}
=\operatorname{diag}(\widehat{\boldsymbol U}_{i,k}
-\widehat{\boldsymbol L}_{i,k})
(\widetilde{\boldsymbol s}_{i,k}-\boldsymbol s_{i,k}^{\rm raw}).
\]
Estimating the root does not by itself create reconstruction error.
The reconstruction error measures only the physical displacement caused by the inward
projection of a within-cell coordinate and is zero when no such projection is
active.
Sample-range endpoints have raw coordinates zero or one and therefore provide
a simple case in which a particle is moved slightly into its terminal cell.
For fixed nonnegative weights $\alpha_i$ that sum to one, let
$\widehat\mu_{Q,n,L}^{\oplus}$ be the continuous Q barycenter of these
auxiliary laws.
Write $\operatorname{diam}_L(\widehat\mu_{Q,n,L}^{\oplus})$ for its largest
depth-$L$ cell diameter and $\operatorname{diam}_0(\widehat\mu_{Q,n,L}^{\oplus})$
for its root diameter.
The two diameters and $e_{i,n,L}^{\rm rec}$ appear only in the bound that
connects the auxiliary continuous barycenter to the root-free output
$\widetilde\mu_{Q,n}$.

For fixed affine weights $\alpha_1,\ldots,\alpha_q$ that sum to one, write
$\bg_i=\Phi(\mu_i)$ and set
\[
\begin{aligned}
\bg_\alpha&=\sum_{i=1}^q\alpha_i\bg_i,\\
\widehat{\bg}_{\alpha,n,L}
&=\iota_L\left(\sum_{i=1}^q
\alpha_i\widehat{\bg}_{i,n,L}\right).
\end{aligned}
\]
Thus $D(\bg_\alpha)$ is the population TL barycenter, while
\eqref{eq:main-practical-tl-output} and
\eqref{eq:main-continuous-tl-output} are its atomic and continuous empirical
counterparts.

Let $N_\sigma$ be the number of ordered positions assigned to leaf
$\sigma\in\{0,1\}^L$, and put $M_L=2^L$.
Balanced integer halving makes every $N_\sigma$ equal to
$\lfloor n/M_L\rfloor$ or $\lceil n/M_L\rceil$.
If $s_{n,L}=n\bmod M_L$, exactly $s_{n,L}$ leaves have the larger count.
By contrast, $\widehat\mu_{{\rm TL},n,L}^{\rm cont}(\alpha)$ assigns mass
$1/M_L$ to the depth-$L$ cell attached to every address.
The auxiliary laws $\widehat\mu_{i,n,L}^{\rm cont}$ and their continuous
Q barycenter have the same equal address masses.
Hence the total variation distance between the empirical and continuous
address-mass vectors is
\begin{equation}
\mathfrak r_{n,L}
:=\frac12\sum_{\sigma\in\{0,1\}^L}
\left|\frac{N_\sigma}{n}-\frac1{M_L}\right|
=\frac{s_{n,L}(M_L-s_{n,L})}{M_Ln}.
\label{eq:main-exact-allocation}
\end{equation}
If $M_L$ divides $n$, every empirical leaf has mass $1/M_L$ and
$\mathfrak r_{n,L}=0$.
For example, $n=10$ and $L=2$ give $M_L=4$, $s_{n,L}=2$, and leaf counts
$(3,3,2,2)$ up to their order.
The empirical leaf masses are $(0.30,0.30,0.20,0.20)$, while the continuous
address masses are $(0.25,0.25,0.25,0.25)$.
Equation~\eqref{eq:main-exact-allocation} gives $\mathfrak r_{n,L}=0.10$, the
mass that cannot be matched within corresponding leaves.

For a completed TL coordinate $\bg$, let $\operatorname{diam}_0(\bg)$ denote
its root diameter and let $\operatorname{diam}_L(\bg)$ denote its largest
depth-$L$ cell diameter.

\begin{theorem}[Empirical HiMAP error decomposition]
\label{thm:main-estimation}
\mbox{}\par\noindent
Fix $q$ laws $\mu_1,\ldots,\mu_q\in\mathcal A^{\rm enc}$ and suppose
$U_{i,1:n}$ are iid from $\mu_i$.
Fix an integer $L\geq1$ with $2^L\leq n$.
All weights below are fixed and known.

\emph{(a) Positive Q barycenter.}
Let $\alpha_i\geq0$ sum to one, and let $\mu_Q^\oplus$ be the corresponding
population Q barycenter.
Then
\begin{align}
W_2(\widetilde\mu_{Q,n},\mu_Q^\oplus)
\leq{}&\sum_i\alpha_i e_{i,n,L}^{\rm rec}
+\operatorname{diam}_L(\widehat\mu_{Q,n,L}^{\oplus})\notag\\
&+\operatorname{diam}_0(\widehat\mu_{Q,n,L}^{\oplus})
\sqrt{\mathfrak r_{n,L}}\notag\\
&+\sum_i\alpha_i
d_{Q,2}(\widehat\mu_{i,n,L}^{\rm cont},\mu_i).
\label{eq:main-q-finite-error}
\end{align}

\emph{(b) Affine TL barycenter.}
Let $\alpha_i\in\mathbb R$ sum to one.
Suppose $\rho^2<\theta<1$.  On the event that the log-width blocks of
$\widehat{\bg}_{\alpha,n,L}$ and $\bg_\alpha$ are bounded above by $B$, the
following inequality holds:
\begin{align}
W_2\{\widetilde\mu_{{\rm TL},n,L}(\alpha),D(\bg_\alpha)\}
\leq{}& C_B\sum_{i=1}^q|\alpha_i|
\|\iota_L(\widehat{\bg}_{i,n,L})-\bg_i\|_{\mathcal G}\notag\\
&+\operatorname{diam}_L(\widehat{\bg}_{\alpha,n,L})
+\operatorname{diam}_0(\widehat{\bg}_{\alpha,n,L})
\sqrt{\mathfrak r_{n,L}}.
\label{eq:main-tl-finite-error}
\end{align}
\end{theorem}

Part (a) separates quadrature reconstruction, terminal resolution, integer
allocation, and continuous Q estimation.
Part (b) separates input encoding, terminal resolution, and integer allocation.
The quantity $e_{i,n,L}^{\rm rec}$ includes any displacement caused by clipping a
within-cell ratio at a sample-range endpoint.
The absolute weights in part (b) reflect the amplification of input encoding
error under signed aggregation.
The tree depth $L$ and internal cloud size $n$ can therefore be chosen
separately from the number of input distributions.
Section~\ref{app:detail-estimation} and
Appendix~\ref{app:proof-main-estimation} give the component results and the
complete proof.

\paragraph{Supporting rates.}
Fix a depth $L$ and suppose that the support widths and density have positive
lower and finite upper bounds, every population split ratio lies in
$[\varepsilon+\delta_L,1-\varepsilon-\delta_L]$ for some $\delta_L>0$, and the
conditional medians are unique and locally identifiable.
These conditions stabilize the empirical cells, keep finite-depth clipping
inactive with probability tending to one, and convert empirical mass error
into split-location error.
Assumption~\ref{ass:fixed-depth} gives the precise constants.
If $2^L\leq n$, then
\begin{equation}
\|\iota_L(\widehat{\bg}_{n,L})-P_L\bg_\mu\|_{\mathcal G}
=O_p(n^{-1/2}).
\label{eq:main-fixed-depth-rate}
\end{equation}
For every fixed law in $\mathcal A^{\rm enc}$, a deterministic, generally
law-dependent sequence $L_n\to\infty$ gives consistency in $d_{\rm TL}$ and, when
$\rho^2<\theta<1$, in $W_2$.
\begin{corollary}[Consistency of empirical HiMAP barycenters]
\label{cor:main-barycenter-consistency}
\mbox{}\par\noindent
Fix $q$ and laws $\mu_1,\ldots,\mu_q\in\mathcal A^{\rm enc}$.
If the fixed weights $\alpha_i$ are nonnegative and sum to one, the root-free
Q output in \eqref{eq:main-practical-q-output} satisfies
\[
W_2(\widetilde\mu_{Q,n},\mu_Q^\oplus)\xrightarrow{p}0.
\]
If instead $\alpha_i\in\mathbb R$ sum to one and $\rho^2<\theta<1$, there is a
deterministic sequence $L_n\to\infty$ with $2^{L_n}/n\to0$ such that
\[
W_2\{\widetilde\mu_{{\rm TL},n,L_n}(\alpha),D(\bg_\alpha)\}
\xrightarrow{p}0.
\]
\end{corollary}

For Q-HiMAP, the common depth sequence is used only in the proof to connect the
root-free particle output to continuous Q paths.
Thus the computed Q output has no depth tuning parameter, whereas TL
computation uses a declared finite depth.
Section~\ref{sec:experiments} evaluates both estimators using the particle
outputs in \eqref{eq:main-practical-q-output} and
\eqref{eq:main-practical-tl-output}.

\section{Numerical Experiments}
\label{sec:experiments}

We first compare the two barycentric geometries on finite problems with known or independently computed solutions.
We then study affine regression, hierarchical aggregation, and repeated prediction.

\subsection{Experimental settings and baselines}
\label{subsec:experiment-settings}

In all experiments, TL-HiMAP takes the coordinatewise sample range as its empirical root and uses the safe-logit buffer $\varepsilon=10^{-3}$ unless stated otherwise.
Q-HiMAP uses nonnegative barycentric weights.
When regression weights are signed, \qaffine{} (empirical) applies them directly to particles matched by their common ranks.

The barycenter and regression experiments use different baselines.
For positive barycenters, we compare HiMAP with the sliced Wasserstein barycenter (SWB) of \citet{BonneelRabinPeyrePfister2015SlicedRadonBarycenters} and the linear barycentric coding model (LBCM) of \citet{WerenskiMalleryAeronMurphy2025Linearized}.
SWB uses $512$ fixed projections.
LBCM uses as its reference the Wasserstein medoid, the input distribution
minimizing the total squared $W_2$ distance to all other inputs.
For Fr\'echet regression, we compare HiMAP with three baselines.
SAW is the signed slice-averaged estimator of \citet{chen2023sliced} and uses $64$ fixed projections.
NPT-FR is the nonparanormal transport regression method of \citet{ParkGaynanova2026NPTFrechet}.
FM is the conditional entropic Wasserstein estimator of \citet{FanMuller2025ConditionalWassersteinBarycenters}.

We compare the two closed-form HiMAP barycenters through the unregularized
empirical $W_2^2$ objective in the ambient Wasserstein geometry.
The primary regression loss is the transport component of entropic optimal
transport with squared Euclidean cost and $\epsilon_{\rm OT}=10^{-3}$.
We average this loss over prediction targets and call it the \emph{mean
predictive Sinkhorn cost}.
In the broad regression study of Section~\ref{subsec:regression-simulations},
we consider $n\in\{256,512\}$ particles per distribution.
At each $n$, every method uses all particles in each training response and
returns an $n$-particle prediction.
Each target is an independent $n$-particle sample drawn directly from the
data-generating distribution.
The loss is computed from the complete $n\times n$ coupling, without particle
ordering or support reduction.
All reported couplings have maximum marginal error below $10^{-5}$.
The particle algorithms do not use the Hilbert weight $\theta$.
For numerical interpretation of the decoder-stability bound, we use the
admissible calibration $\theta=(1+\rho^2)/2$ described in
Appendix~\ref{app:detail-stability}.

\FloatBarrier

\subsection{Calibration of the two barycentric geometries}
\label{subsec:positive-barycenters}

\paragraph{Calibration against certified $W_2$ barycenters.}

We use three controlled multimodal examples to compare the induced geometries with independently verified $W_2$ solutions.
The checkerboard, two-spirals, and pinwheel problems each contain three equally weighted, $64$-atom distributions arranged in sixteen macro modes.
The examples vary the global topology while sharing a prescribed three-marginal correspondence.
Averaging matched atoms gives a closed-form candidate target.
A finite Kantorovich dual certificate over all $64^3$ triples verifies $W_2$ optimality, with largest primal--dual gap below $5\times10^{-19}$.

\begin{table}[!t]
\centering
\caption{Percentage excess of the unregularized $W_2$ barycenter objective
relative to independently certified positive-weight targets.
All inputs and outputs contain $64$ equally weighted particles.
Dark orange and light orange cells mark the best and second-best values within each
column.}
\label{tab:positive-barycenters}
\small
\begin{tabular}{lrrr}
\toprule
Method & Checkerboard & Two spirals & Pinwheel \\
\midrule
Q-HiMAP & \bestscore{$0.00$} & \bestscore{$0.00$} & \bestscore{$0.00$} \\
TL-HiMAP & \secondscore{$19.45$} & \secondscore{$12.03$} & \secondscore{$10.56$} \\
SWB (512 projections) & $24.75$ & $52.58$ & $29.14$ \\
LBCM & $45.33$ & $47.53$ & $50.12$ \\
Grid Sinkhorn & $751.71$ & $1917.70$ & $2753.14$ \\
\bottomrule
\end{tabular}
\end{table}

\begin{figure}[p]
\centering
\includegraphics[width=0.98\textwidth]{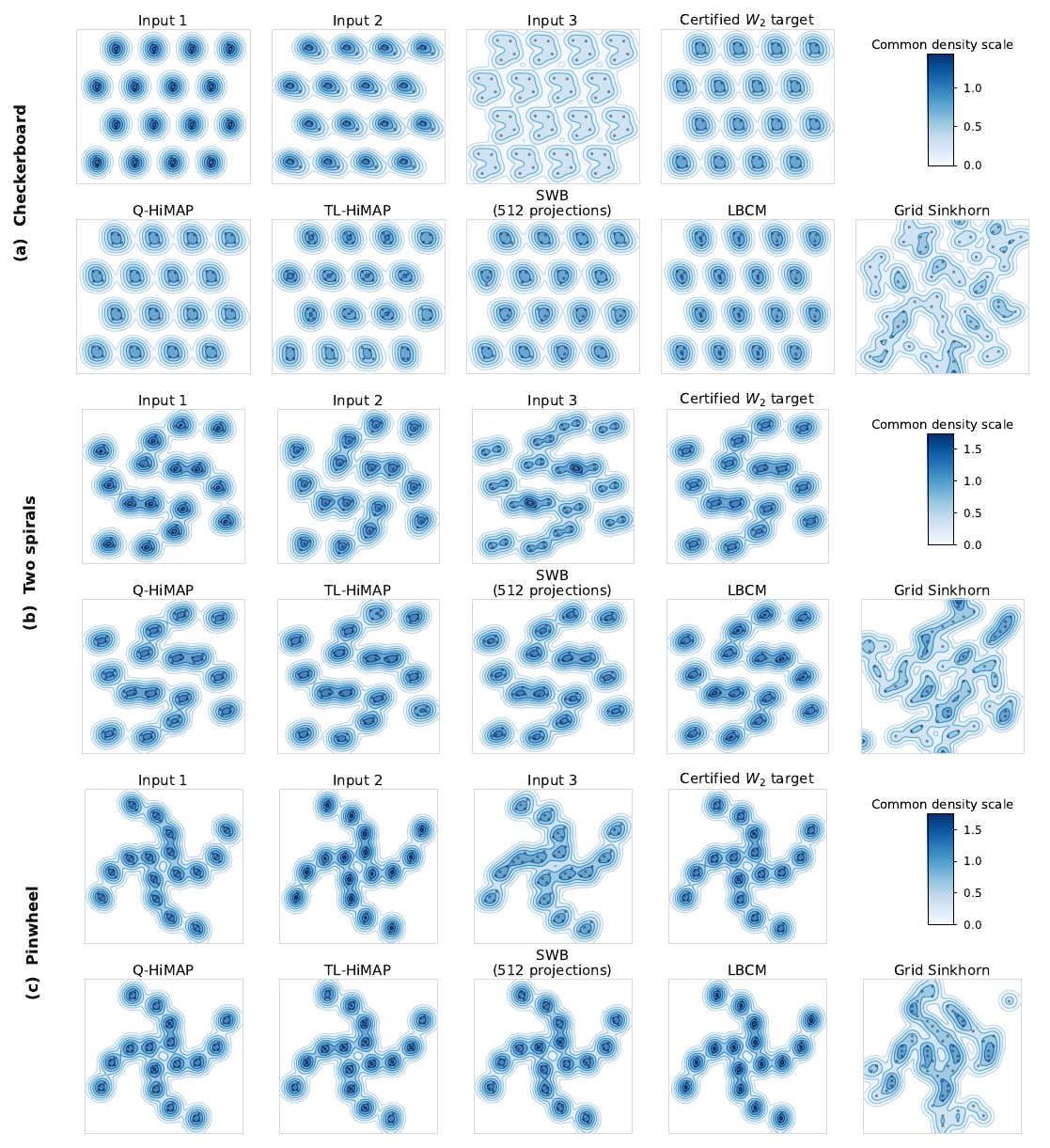}
\caption{Complete certified positive-weight barycenter comparison.
Panels (a)--(c) show the checkerboard, two spirals, and pinwheel geometries.
Within each panel block, the upper row contains the three inputs and the
certified $W_2$ target.
The lower row contains Q-HiMAP, TL-HiMAP, SWB with $512$ projections, LBCM,
and Grid Sinkhorn.
Bounds, bandwidth, contour levels, and the absolute density scale are common
within each topology.
Every distribution contains $64$ equally weighted particles.}
\label{fig:positive-barycenters-complete}
\end{figure}

Table~\ref{tab:positive-barycenters} therefore measures agreement with a known $W_2$ solution under three distinct multimodal layouts.
Q-HiMAP recovers all three certified targets to numerical precision because the optimal coupling follows the common mass address.
Figure~\ref{fig:positive-barycenters-complete} shows the three inputs, the certified target, and every barycenter estimate for each topology.

Across the three layouts, Q-HiMAP matches the certified target, while TL-HiMAP preserves the macro-mode arrangement and has lower $W_2$ excess objective than SWB, LBCM, and Grid Sinkhorn.
Thus Q-HiMAP and TL-HiMAP both produce barycenters close to the certified $W_2$
solution in these examples, although only Q-HiMAP directly averages the
prescribed correspondence.

\subsection{Broad distribution regression}
\label{subsec:regression-simulations}

\paragraph{Data-generating family.}
Both regression experiments use the same bounded multimodal family.
Let $\boldsymbol C_1,\ldots,\boldsymbol C_K$ be replicate-specific component
centers, let $p_k$ and $\tau_k$ denote their base masses and mass trend scores,
and let $\bv$ be a unit direction.
For response dimension $d$ and $\gamma\in\{0,1\}$, the target distribution
$\mu_\gamma(x)$ is the law of
\begin{equation}
\begin{aligned}
U_\gamma(x)&=A_\gamma(x)\{\boldsymbol C_{J_\gamma(x)}+\boldsymbol E\}+\bb_\gamma(x),\\
\Pr\{J_\gamma(x)=k\}
&\propto p_k\exp\{0.65\gamma\sin(2\pi x)\tau_k\},\\
A_\gamma(x)&=R_{12}\{0.55\gamma\sin(\pi x)\}D_\gamma(x),\\
D_\gamma(x)&=\operatorname{diag}_{j\leq d}
 \left[\exp\{0.22\gamma\sin(2\pi x+\phi_j)\}\right],\\
\bb_\gamma(x)&=(1-\gamma)0.65x\bv
+\gamma\{0.28\sin(2\pi x)\bv+0.18\cos(2\pi x)\boldsymbol e_2\}.
\end{aligned}
\label{eq:regression-simulation-family}
\end{equation}
The noise vector $\boldsymbol E=(E_1,\ldots,E_d)^\top$ is independent of $J_\gamma(x)$.
Its coordinates are independent and uniformly distributed on $[-0.11\sqrt{3},0.11\sqrt{3}]$.
The matrix $R_{12}(a)$ rotates the first two coordinates by angle $a$, and $\boldsymbol e_2$ is the second coordinate vector.
Appendix~\ref{sec:experimental-setup} gives all constants.
At $\gamma=0$, component masses and shapes are fixed.
The full distribution follows the affine path $0.65x\bv$, which is fitted with
global Fr\'echet weights.
At $\gamma=1$, component masses, coordinate scales, orientation, and location
vary nonlinearly with $x$.
Local-linear Fr\'echet weights with bandwidth $0.22$ fit this nonlinear
trajectory.

Each experiment contains $31$ training distributions, and we consider
$n\in\{256,512\}$ particles per response.
At each of the nine target values and each $n$, we draw a new independent
$n$-particle sample directly from $\mu_\gamma(x)$.
The evaluation covers $d\in\{2,5,10,20\}$ over $30$ replicates.
Every method uses all training particles and returns $n$ equally weighted
particles.
The mean predictive Sinkhorn cost uses the complete $n\times n$ coupling.
TL-HiMAP uses depths eight and nine for $n=256$ and $512$, respectively, so
each terminal cell contains one particle.

\begin{figure}[!t]
\centering
\includegraphics[width=0.80\textwidth]{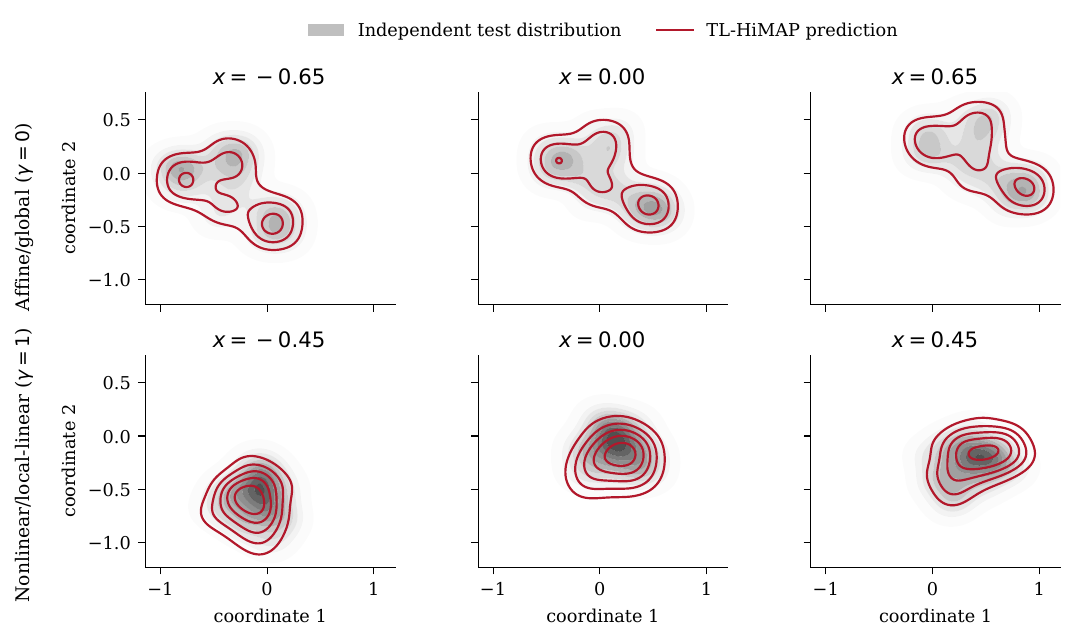}
\caption{Representative trajectories and predictions in dimension two.
The columns show the first, middle, and last prediction targets from repeat
zero.
Gray shading denotes the independent target density, and red contours denote
the TL-HiMAP prediction.
The displayed target and prediction contain $256$ equally weighted particles.
Each row uses common bounds, bandwidth, contour levels, and absolute density scale.}
\label{fig:broad-regression-trajectories}
\end{figure}

Figure~\ref{fig:broad-regression-trajectories} shows the first, middle, and
last target in the first two-dimensional replicate.
The global trajectory in the upper row preserves masses and shapes while
translating every mode along one direction, and its two displayed endpoints
lie outside the training interval $[-0.5,0.5]$.
By contrast, the local trajectory in the lower row changes component masses,
anisotropy, orientation, and location nonlinearly.

\paragraph{Results.}

\begin{table}[!t]
\centering
\caption{Mean predictive Sinkhorn cost at two particle counts, with standard
deviations over $30$ replicates in parentheses. Each method uses all particles
in the $31$ training responses and returns the same number of particles as the
independently generated target sample. The cost uses the complete $n\times n$
coupling. Smaller values indicate better prediction. FM is evaluated in
dimension two. Dark orange and light orange mark the best and second-best
values within each row.}
\label{tab:broad-regression}
\scriptsize
\setlength{\tabcolsep}{3.2pt}
\begin{tabular}{llcrrrrr}
\toprule
Trajectory/fit & $d$ & $n$ & TL-HiMAP & Q-affine (emp.)
& SAW (64) & NPT-FR & FM \\
\midrule
$\gamma=0$, global & 2 & 256
& \secondscore{$0.0038\,(0.0009)$}
& \secondscore{$0.0038\,(0.0009)$}
& \bestscore{$0.0036\,(0.0009)$}
& $0.0090\,(0.0047)$ & $0.0086\,(0.0016)$ \\
& & 512
& $0.0025\,(0.0006)$
& \secondscore{$0.0024\,(0.0006)$}
& \bestscore{$0.0023\,(0.0005)$}
& $0.0079\,(0.0048)$ & $0.0073\,(0.0016)$ \\
\addlinespace[1pt]
& 5 & 256
& \secondscore{$0.0246\,(0.0012)$}
& $0.0253\,(0.0014)$
& \bestscore{$0.0245\,(0.0013)$}
& $0.0377\,(0.0054)$ & -- \\
& & 512
& \secondscore{$0.0184\,(0.0007)$}
& $0.0186\,(0.0009)$
& \bestscore{$0.0179\,(0.0007)$}
& $0.0313\,(0.0051)$ & -- \\
\addlinespace[1pt]
& 10 & 256
& \bestscore{$0.0796\,(0.0016)$}
& \secondscore{$0.0809\,(0.0022)$}
& $0.0894\,(0.0017)$
& $0.1080\,(0.0052)$ & -- \\
& & 512
& \bestscore{$0.0681\,(0.0010)$}
& \secondscore{$0.0686\,(0.0015)$}
& $0.0752\,(0.0010)$
& $0.0944\,(0.0048)$ & -- \\
\addlinespace[1pt]
& 20 & 256
& \bestscore{$0.2088\,(0.0030)$}
& \secondscore{$0.2115\,(0.0051)$}
& $0.2570\,(0.0021)$
& $0.2795\,(0.0038)$ & -- \\
& & 512
& \bestscore{$0.1913\,(0.0027)$}
& \secondscore{$0.1965\,(0.0047)$}
& $0.2327\,(0.0013)$
& $0.2572\,(0.0036)$ & -- \\
\midrule
$\gamma=1$, local-linear & 2 & 256
& \secondscore{$0.0164\,(0.0051)$}
& \secondscore{$0.0164\,(0.0050)$}
& \bestscore{$0.0162\,(0.0050)$}
& $0.0201\,(0.0068)$ & $0.0230\,(0.0070)$ \\
& & 512
& \secondscore{$0.0154\,(0.0044)$}
& \secondscore{$0.0154\,(0.0045)$}
& \bestscore{$0.0152\,(0.0044)$}
& $0.0193\,(0.0063)$ & $0.0221\,(0.0065)$ \\
\addlinespace[1pt]
& 5 & 256
& \bestscore{$0.0387\,(0.0032)$}
& $0.0398\,(0.0034)$
& \secondscore{$0.0393\,(0.0031)$}
& $0.0505\,(0.0049)$ & -- \\
& & 512
& $0.0332\,(0.0029)$
& \secondscore{$0.0331\,(0.0029)$}
& \bestscore{$0.0329\,(0.0028)$}
& $0.0440\,(0.0044)$ & -- \\
\addlinespace[1pt]
& 10 & 256
& \bestscore{$0.0938\,(0.0027)$}
& \secondscore{$0.0964\,(0.0034)$}
& $0.1064\,(0.0025)$
& $0.1192\,(0.0047)$ & -- \\
& & 512
& \bestscore{$0.0833\,(0.0021)$}
& \secondscore{$0.0839\,(0.0026)$}
& $0.0928\,(0.0024)$
& $0.1056\,(0.0046)$ & -- \\
\addlinespace[1pt]
& 20 & 256
& \bestscore{$0.2259\,(0.0041)$}
& \secondscore{$0.2349\,(0.0060)$}
& $0.2758\,(0.0032)$
& $0.2903\,(0.0045)$ & -- \\
& & 512
& \bestscore{$0.2109\,(0.0037)$}
& \secondscore{$0.2200\,(0.0056)$}
& $0.2526\,(0.0031)$
& $0.2679\,(0.0036)$ & -- \\
\bottomrule
\end{tabular}
\end{table}

Table~\ref{tab:broad-regression} reports absolute prediction errors.
At both particle counts, TL-HiMAP has the lowest mean loss at $d=10$ and $20$
for both trajectories.
SAW is lower at $d=2$, while the three mass-address and slice-based procedures
are close at $d=5$.
At $d=10$ and $20$, TL-HiMAP reduces mean loss relative to SAW by
$9.4\%$--$18.7\%$ and relative to NPT-FR by $21.0\%$--$27.7\%$ across the two
particle counts.
For each replicate, we first average the losses over its nine prediction
targets and then compute the relative difference
$100(\overline L_{\mathrm{TL}}-\overline L_{\mathrm{base}})/
\overline L_{\mathrm{base}}$.
At both particle counts, the $95\%$ bootstrap intervals obtained by resampling the $30$ replicate-level differences are entirely below zero at $d=10$ and $20$.
Doubling the particle count lowers the TL-HiMAP mean loss by
$8.4\%$--$35.2\%$ for the global trajectory and $5.8\%$--$14.4\%$ for the
local-linear trajectory.

The global targets have median minimum weight $-0.0267$, median weight
$\ell^1$ norm $1.21$, and a maximum negative fraction of $35.5\%$.
Thus the global experiment directly exercises signed affine weights.

\begin{figure}[t]
\centering
\includegraphics[width=0.53\textwidth]{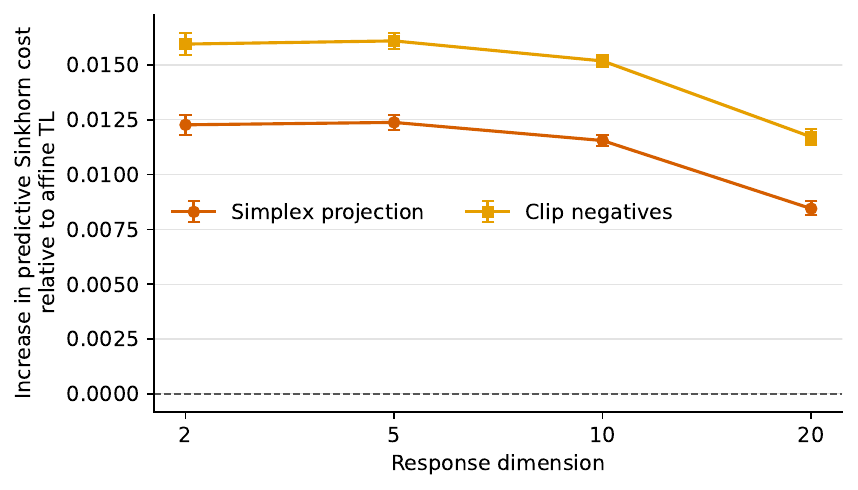}
\caption{Effect of replacing global affine Fr\'echet weights by simplex
projection or negative clipping.
Within each replicate, we first average loss over the nine prediction targets and subtract the corresponding mean loss under the original affine weights.
Points are the mean of these $30$ replicate-level differences, and bars are $95\%$ bootstrap intervals.}
\label{fig:global-weight-repair}
\end{figure}

We next replace each global affine weight vector by either its Euclidean
projection onto the probability simplex or negative clipping followed by
renormalization.
Every encoded response and tuning choice remains unchanged.
Simplex projection increases mean loss by
$324.1\%$, $50.3\%$, $14.5\%$, and $4.0\%$ in dimensions
$2,5,10,$ and $20$.
Negative clipping increases it by $421.4\%$, $65.4\%$, $19.1\%$, and
$5.6\%$.
For every dimension and both repairs, the $95\%$ bootstrap interval for the loss increase relative to the original affine weights lies above zero.
Figure~\ref{fig:global-weight-repair} shows these loss increases.

\FloatBarrier

\subsection{Composition and repeated prediction}
\label{subsec:timing}

\paragraph{Direct versus two-stage prediction.}

We compare a direct aggregation of eight inputs with a two-stage aggregation of the same inputs and weights.
The predictor values are $x_j=-0.42+0.84(j-1)/7$ for $j=1,\ldots,8$.
Each input contains $257$ particles from a five-component bivariate mixture whose centers, masses, and axis scales vary smoothly with $x$.
Appendix~\ref{sec:composition-settings} gives the complete generator.
The translation control draws one particle cloud and shifts it by $\bb(x_j)=\{0.45x_j,0.18\sin(\pi x_j)\}^\top$.
It checks translation equivariance and is not used to rank the methods.
The independent-mixture setting instead draws a new cloud at every $x_j$, allowing both the mixture parameters and sampled particles to vary across inputs.

The weights are
\[
\boldsymbol w(s)=\frac18\boldsymbol{1}_8
+s(1,-1,0.6,-0.6,-1,1,-0.6,0.6)^\top,
\qquad s\in\{0,0.18,0.35\}.
\]
These choices give uniform, moderate affine, and strong affine weighting, with
minimum weights $0.125$, $-0.055$, and $-0.225$.
The two-stage calculation first aggregates the pairs $(1,2)$, $(3,4)$, $(5,6)$, and $(7,8)$ and then combines the four resulting distributions.
For pair $A_r=\{2r-1,2r\}$ with total weight $W_r=\sum_{j\in A_r}w_j(s)$, the first stage uses weights $w_j(s)/W_r$ and the final stage uses $W_r$.
Their product is the original weight $w_j(s)$.
Let $\widehat\nu_{M,s}^{\mathrm{direct}}$ and $\widehat\nu_{M,s}^{\mathrm{two}}$ denote the direct and two-stage outputs of method $M$.
Both contain $257$ equally weighted particles, and Table~\ref{tab:affine-composition} reports the mean of $E_{M,s}=W_2^2(\widehat\nu_{M,s}^{\mathrm{direct}},\widehat\nu_{M,s}^{\mathrm{two}})$ over $30$ replicates.
The empirical transport problem uses the full $257\times257$ cost matrix.

\begin{table}[!htbp]
\centering
\caption{Mean empirical $W_2^2$ between direct and two-stage Fr\'echet
predictions.
Each entry averages $30$ replicates under the two-stage aggregation.
Smaller values indicate closer agreement under composition.
Dark orange and light orange cells mark the best and second-best values within
each independent-mixture column; the translation control is not ranked.}
\label{tab:affine-composition}
\scriptsize
\begin{adjustbox}{max width=\textwidth}
\begin{tabular}{lrrrrrr}
\toprule
& \multicolumn{3}{c}{Translation control}
& \multicolumn{3}{c}{Independent mixture} \\
\cmidrule(lr){2-4}\cmidrule(lr){5-7}
Method & $s=0$ & $s=0.18$ & $s=0.35$
& $s=0$ & $s=0.18$ & $s=0.35$ \\
\midrule
TL-HiMAP
& $5.54\times10^{-10}$ & $5.54\times10^{-10}$ & $5.54\times10^{-10}$
& \secondscore{$6.51\times10^{-5}$} & \bestscore{$6.86\times10^{-5}$} & \bestscore{$5.16\times10^{-4}$} \\
Q-affine (empirical)
& $6.78\times10^{-18}$ & $8.19\times10^{-18}$ & $8.87\times10^{-18}$
& \bestscore{$9.35\times10^{-18}$} & $6.34\times10^{-4}$ & $4.51\times10^{-3}$ \\
SAW (64 projections)
& $8.57\times10^{-18}$ & $1.11\times10^{-17}$ & $9.71\times10^{-18}$
& $6.93\times10^{-4}$ & $5.75\times10^{-4}$ & $1.88\times10^{-3}$ \\
NPT-FR
& $8.22\times10^{-4}$ & $8.22\times10^{-4}$ & $8.22\times10^{-4}$
& $2.53\times10^{-4}$ & \secondscore{$3.26\times10^{-4}$} & \secondscore{$8.80\times10^{-4}$} \\
\bottomrule
\end{tabular}
\end{adjustbox}
\end{table}

Before particle reconstruction, the affine TL coordinate average is algebraically compositional.
Across the $30$ replicates, both data-generating models, and all three values of $s$, the largest Euclidean difference between the direct and two-stage TL coordinate vectors is $2.61\times10^{-15}$.
The empirical outputs also include decoding and re-encoding at the intermediate stage, so their composition errors depend on the setting.
In the translation control, Q-affine and SAW agree up to numerical precision, while TL-HiMAP has a small finite reconstruction discrepancy; these columns are not ranked.
For the independent mixtures, Q-affine is exact at $s=0$, whereas TL-HiMAP has the smallest error at the two signed weight levels $s=0.18$ and $s=0.35$.

\paragraph{Repeated prediction.}

\begin{figure}[!t]
\centering
\includegraphics[width=0.79\textwidth]{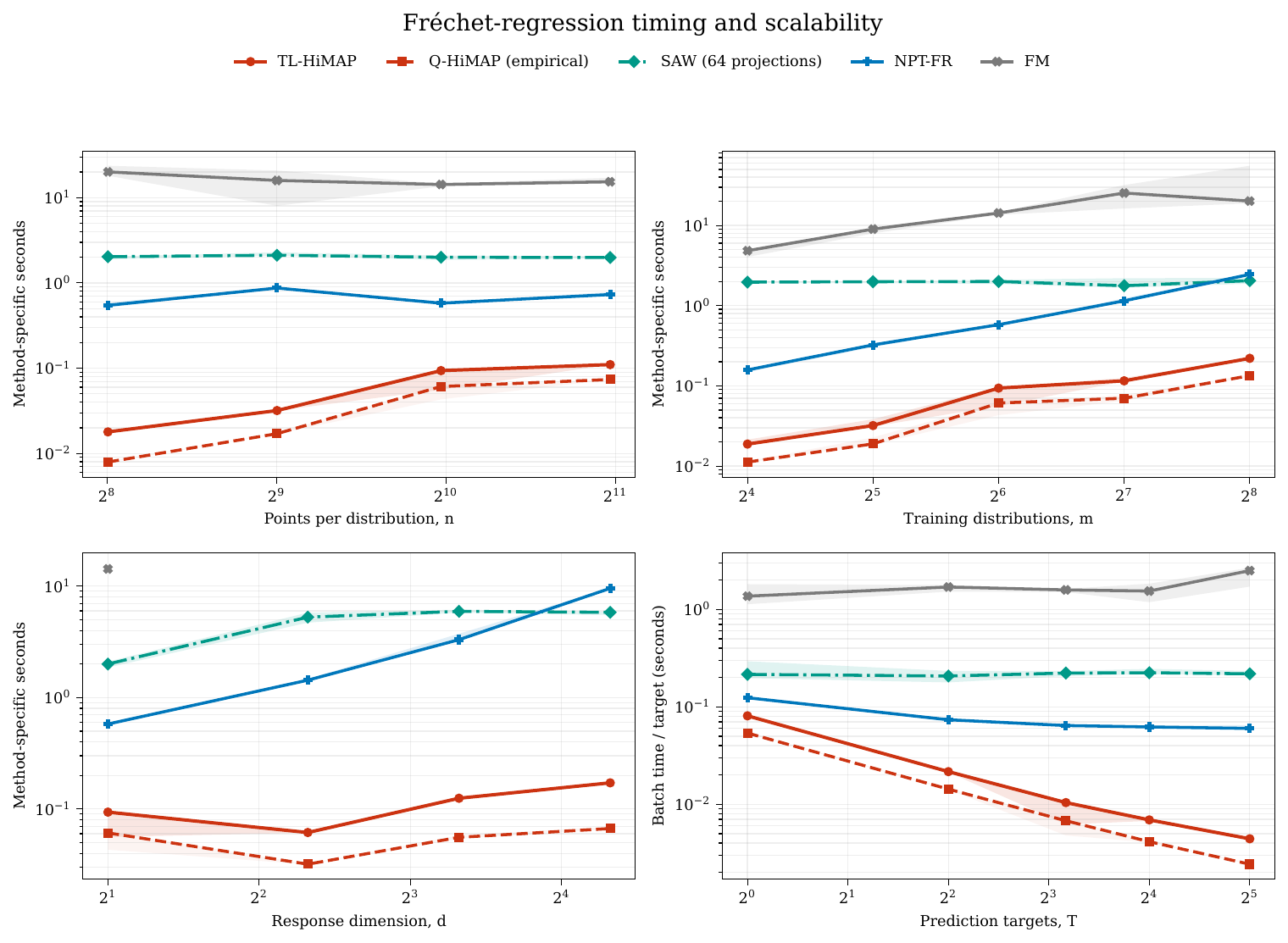}
\caption{Single-core method-specific fit-and-predict time as particles,
training distributions, response dimension, and prediction targets vary.
The final panel divides the total fit-and-predict time for each batch by its
number of targets.
It therefore reports amortized time per target rather than total batch time.
Lines are medians and bands are interquartile ranges.}
\label{fig:timing}
\end{figure}

An isolated single-core benchmark varies the number of distributions,
particles, response dimensions, and prediction targets.
The timer includes construction of each reusable response representation,
affine aggregation, and decoding for the full target batch.
Timing excludes data generation, the common regression-weight calculation, loss
evaluation, plotting, and input/output.
Runs use one logical core of an Intel Xeon Platinum 8280 processor, one
numerical library thread, double precision, Python 3.9.12, and NumPy 1.26.4.
At $m=64$, $n=1003$, $d=2$, and nine targets, median method-specific
fit-and-predict times are $0.0610$ seconds for Q-affine (empirical), $0.0937$
for TL-HiMAP, $0.5785$ for NPT-FR, $2.0003$ for SAW with $64$ projections,
and $14.2705$ for FM.
When the target count increases from one to $32$, the TL-HiMAP total rises
from $0.0808$ to $0.1416$ seconds.
The TL-HiMAP median time per target falls from $0.0808$ to $0.00443$ seconds.
This $18$-fold reduction in the final panel of Figure~\ref{fig:timing} reflects
the reuse of each response representation across targets.
Table~\ref{tab:timing-decomposition} separates reusable construction from the
nine-target prediction batch at the same reference cell.

The next section applies the same regression estimators and evaluation
criterion to participant-level activity distributions from two NHANES cycles.

\section{NHANES Activity-Distribution Regression}
\label{sec:nhanes}

We use public-use wrist accelerometer records from the National Health and
Nutrition Examination Survey to study participant-level activity
distributions \citep{CDC_NHANES2011_2014}.
The 2011--2012 cycle determines the model specification and tuning parameters.
We then apply the selected model specification and tuning parameters unchanged
to the independent 2013--2014 cycle.
Both cycles use the same eligibility criteria, response definition, fold
assignment, loss, and profile summaries.

\subsection{Data and evaluation}

Physical activity varies within a participant across the day, so a single daily mean can hide differences between sustained and intermittent movement.
We retain this within-participant heterogeneity by representing each person as a distribution of local activity summaries.
The statistical unit is a participant aged $18$--$79$ years with complete demographic, clinical, survey-design, and accelerometer information.
For each participant, four central wear days are divided into overlapping six-hour windows with a one-hour stride, which provides dense coverage of within-day changes.
Each valid window contributes $(\log(1+\text{mean activity}),\log(1+\text{coefficient of variation}))$.
The first coordinate measures activity level, while the second measures variation relative to that level.
The empirical distribution of these vectors is the response.
It distinguishes participants with similar overall activity but different proportions of low- and high-activity periods or different local variability.
Because the overlapping windows are dependent, participants rather than windows define the folds and units used for inference.

These criteria yield $4091$ eligible participants in 2011--2012 and $4262$ in
2013--2014.
For each training participant, an outcome-independent deterministic ordering
based on the survey cycle, participant identifier, wear day, and window start
time selects $32$ valid windows without replacement.
All eligible windows of a participant assigned to a test fold form the
evaluation distribution, and none enters model fitting.
Five subject-disjoint folds balance sex, age quintile, and glycated hemoglobin
(HbA1c) category within each cycle.
HbA1c summarizes average blood glucose over the preceding two to three months
\citep{CDC_A1C2024}.
Within each training fold, age, body mass index, and HbA1c are standardized
using survey-weighted means and standard deviations.
The predictor vector contains these three standardized variables and a female
indicator.
Writing $A^*$ for standardized age, it also contains $(A^*)^2$ minus its
survey-weighted training-fold mean to represent a quadratic age association.
Response scaling also uses the training fold only.
NHANES assigns each participant a Mobile Examination Center (MEC) examination
weight that adjusts for unequal selection probabilities, examination
nonresponse, and poststratification \citep{CDC_NHANESWeighting}.
Let $\omega_i$ denote this weight after normalization within a training fold.
We use these normalized values to compute the predictor mean
$\overline X_\omega$ and covariance $\widehat\Sigma_\omega$.
For target $\bx$, all methods use
\[
\alpha_i(\bx)\propto\omega_i\{1+(X_i-\overline X_\omega)^\top
\widehat\Sigma_\omega^+(\bx-\overline X_\omega)\},
\]
with the proportionality constant chosen so that $\sum_i\alpha_i(\bx)=1$.
Appendix~\ref{sec:nhanes-profiles} gives the full definition.

All methods receive the same training distributions and affine weights.
TL-HiMAP, \qaffine{} (empirical), SAW with $64$ projections, and NPT-FR cover
every eligible participant.
The primary loss is the Sinkhorn transport cost between each $32$-particle
prediction and the full evaluation distribution, with
$\epsilon_{\rm OT}=10^{-3}$.
Design-based intervals use $10{,}000$ bootstrap samples of masked primary
sampling units within strata and the survey weights.

The global weights frequently take negative values, with median negative-weight
fractions of $31.0\%$ and $30.9\%$ in the 2011--2012 and 2013--2014 cycles,
respectively.
Their median $\ell^1$ norms are $1.72$ and $1.71$, and their largest norms are
$6.50$ and $5.34$.
Every reported method returns a valid prediction.
The largest Sinkhorn marginal residual is $9.98\times10^{-6}$.

\subsection{Results}

TL-HiMAP has $4.0\%$ lower loss than NPT-FR in 2011--2012 and $3.0\%$ lower
loss in 2013--2014.
The intervals for the TL-to-\qaffine{} ratio include one in both cycles.
TL-HiMAP has $4.9\%$ and $5.8\%$ higher primary loss than SAW.
TL-HiMAP also has lower loss than FM in both cycles.
Table~\ref{tab:nhanes} reports TL-to-baseline loss ratios computed on the same
test-fold participants.
Figure~\ref{fig:nhanes} in the supplement gives the corresponding forest plot.

\begin{table}[!htbp]
\centering
\caption{Survey-weighted ratio of TL-HiMAP predictive Sinkhorn loss to each
baseline.
Values below one favor TL-HiMAP, and parentheses give design-based $95\%$ intervals obtained by resampling masked primary sampling units within strata.
Each row compares TL-HiMAP and one baseline on the same set of test participants, so the rows are not ranked against one another.}
\label{tab:nhanes}
\small
\begin{tabular}{lcc}
\toprule
Baseline & 2011--2012 & 2013--2014 \\
\midrule
NPT-FR & $0.960\;(0.952,0.966)$ & $0.970\;(0.965,0.975)$ \\
SAW (64 projections) & $1.049\;(1.042,1.056)$ & $1.058\;(1.053,1.064)$ \\
\qaffine{} (empirical) & $0.994\;(0.985,1.001)$ & $1.002\;(0.997,1.007)$ \\
FM & $0.788\;(0.730,0.851)$ & $0.924\;(0.849,0.995)$ \\
\bottomrule
\end{tabular}
\end{table}
\FloatBarrier

\begin{figure}[!htbp]
\centering
\includegraphics[width=0.80\textwidth]{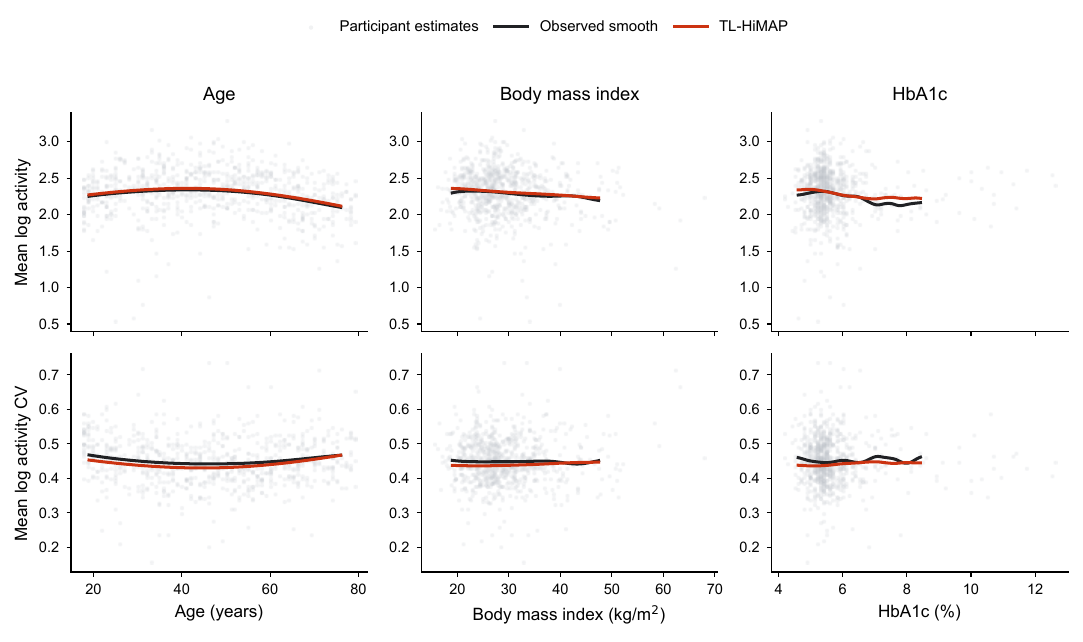}
\caption{Distributional regression profiles in the independent 2013--2014 NHANES cycle.
Grey points summarize the full evaluation distributions, while the black and red curves apply the same survey-weighted local-linear smoother to the observed summaries and TL-HiMAP predictions.
Predictor grids and bandwidths are fixed across panels.}
\label{fig:nhanes-profiles-main}
\end{figure}

Figure~\ref{fig:nhanes-profiles-main} compares mean log activity and mean log coefficient of variation across age, body mass index, and HbA1c.
TL-HiMAP reproduces the shallow inverted-U age profile and the decline in mean log activity with body mass index, while mean log coefficient of variation changes less.
Section~\ref{sec:nhanes-profiles} gives bivariate contours and fitted latent-dependence profiles for the four methods evaluated for every eligible participant.

The age and body mass index profiles recur in both cycles.
The observed age and body mass index associations are descriptive rather than
causal.

\FloatBarrier

\section{Discussion}
\label{sec:discussion}

HiMAP provides a common probability address for barycenters and regression
with multivariate distributions.
Q-HiMAP defines positive barycenters in an induced spatial geometry, whereas
TL-HiMAP uses affine Hilbert coordinates for barycenters with signed weights
and for repeated composition.
These are HiMAP-induced barycenters and need not minimize the conventional
$W_2$ Fr\'echet objective.
Nevertheless, they are competitive under external $W_2$ evaluation in our
numerical studies.
The theory establishes exact mass coding and inversion, closure and transport
stability, deterministic empirical error decompositions, and $W_2$-density of
the union of buffered TL model classes.
The current methods use a fixed coordinate schedule, and TL-HiMAP also uses
sample-range roots.
These choices can make performance sensitive to coordinate orientation or
extreme observations.
Future work may develop adaptive balanced schedules and robust root estimators,
and study gradient flows or other iterative operations in TL coordinates.

\bibliography{ref}

\end{document}